\documentclass[twocolumn,showpacs,preprintnumbers,amsmath,amssymb,floatfix,superscriptaddress]{revtex4-1}
\usepackage{graphicx}
\usepackage{amsmath,amsfonts,amssymb}
\usepackage{graphicx}
\usepackage{color}
\usepackage[stable]{footmisc}

\usepackage[normalem]{ulem}
\usepackage{todonotes}
\usepackage{bbold}
\usepackage[bookmarks, 
            breaklinks, 
            pdfstartview=FitH, 
            pdftitle={},
            pdfauthor={},
            pdfkeywords={}
            ]{hyperref}
\usepackage{ulem}
\usepackage{floatflt}
\usepackage{float}

\def\beq{\begin{equation}}
\def\eeq{\end{equation}}

\newcommand{\bra}[1]{\langle #1|}
\newcommand{\ket}[1]{|#1\rangle}

\newcommand{\rf}[0]{\Omega_{\rm rf}}

\definecolor{ao}{rgb}{0.0, 0.5, 0.0}

\definecolor{gb}{rgb}{0.0, 0.5, 0.5}

\newcommand{\pp}[0]{\hat{\mathbf p}}
\newcommand{\rr}[0]{\mathbf r}
\newcommand{\xx}[0]{\mathbf x}
\newcommand{\RR}[0]{\mathbf R}

\newcommand{\bs}[0]{\hat{\mathbf S}}

\newcommand{\affA}{Institut f{\"u}r Physik, Johannes Gutenberg-Universit{\"a}t Mainz, D-55099 Mainz, Germany}

\newcommand{\affB}{Institute for Quantum Optics and Quantum Information of the Austrian Academy of Sciences, A-6020 Innsbruck, Austria}
\newcommand{\affC}{Institute for Theoretical Physics, University of Innsbruck, A-6020 Innsbruck, Austria}
\newcommand{\affD}{Zentrum f\"ur Optische Quantentechnologien and The Hamburg Centre for Ultrafast Imaging, Universit\"at Hamburg,  Luruper Chaussee 149, D-22761 Hamburg, Germany}

\begin{document}

\title{Controlled long-range interactions between Rydberg atoms and ions}

\author{T.~Secker}\thanks{These authors contributed equally and share first authorship.}\affiliation{\affA}
\author{R.~Gerritsma}\thanks{These authors contributed equally and share first authorship.}\affiliation{\affA}\email{rene.gerritsma@uni-mainz.de}
\author{A.~W. Glaetzle}\affiliation{\affB}\affiliation{\affC}
\author{A.~Negretti}\affiliation{\affD}

\date{\today}

\begin{abstract}
We theoretically investigate trapped ions interacting with atoms that are coupled to Rydberg states. The strong polarizabilities of the Rydberg levels increases the interaction strength between atoms and ions by many orders of magnitude, as compared to the case of ground state atoms, and may be mediated over micrometers. We calculate that such interactions can be used to generate entanglement between an atom and the motion or internal state of an ion. Furthermore, the ion could be used as a bus for mediating spin-spin interactions between atomic spins in analogy to much employed techniques in ion trap quantum simulation. The proposed scheme comes with attractive features as it maps the benefits of the trapped ion quantum system onto the atomic one without obviously impeding its intrinsic scalability. No ground state cooling of the ion or atom is required and the setup allows for full dynamical control. Moreover, the scheme is to a large extent immune to the micromotion of the ion. Our findings are of interest for developing hybrid quantum information platforms and for implementing quantum simulations of solid state physics.
\end{abstract}


\maketitle

\section{Introduction}

Given the highly successful use of trapped ions and ultracold atoms in studying quantum physics, it is of considerable interest to couple these two systems on the quantum level~\cite{Grier:2009,Zipkes:2010,Schmid:2010,Zipkes:2010b,Harter:2013} . The main features of each may thus be combined to break new ground in studying quantum many-body physics~\cite{Bissbort:2013}. Proposals aimed at quantum information processing ~\cite{Doerk:2010} and the generation of entangled atom-ion systems have been put forward recently~\cite{Gerritsma:2012,Joger:2014,Schurer:2016}, which point the way towards such hybrid atom-ion quantum systems. However, the time-dependent trapping field of the ions in a Paul trap poses a significant obstacle for these ideas as it limits attainable temperatures in interacting atom-ion systems~\cite{Zipkes:2010,Schmid:2010,Nguyen:2012,Cetina:2012, Krych:2013,Chen:2014,Weckesser:2015}. This effect stems from the fast micromotion of ions trapped in radio frequency traps which may cause coupling to high energy states when collisions with atoms occur. Additional optical potentials for the atoms that prevent the atoms from colliding with the ions, while still allowing significant interactions, are quite challenging to implement. This is because the atom-ion interaction typically takes place on the 100~nm scale and has a steep $-1/R^4$ character, with $R$ the distance between the atom and ion, making it very hard to optically resolve. These issues put severe restraints on proposed schemes for generating entanglement between atoms and ions~\cite{Nguyen:2012,Joger:2014}, such as employing controlled collisions with state-dependent scattering length~\cite{Doerk:2010} and coupling single ions to atomic Josephson junctions~\cite{Gerritsma:2012,Joger:2014}, where  the atoms have to be brought very close to the ion. Other experimental approaches such as octupole traps with nearly field-free regions~\cite{Deiglmayr:2012} and optical trapping of the ions~\cite{Enderlein:2012}, seriously reduce the merits of the trapped ion platform such as long lifetimes and localisation of individual ions.
\begin{figure}[tb]
\centering
\includegraphics[width=0.85 \columnwidth]{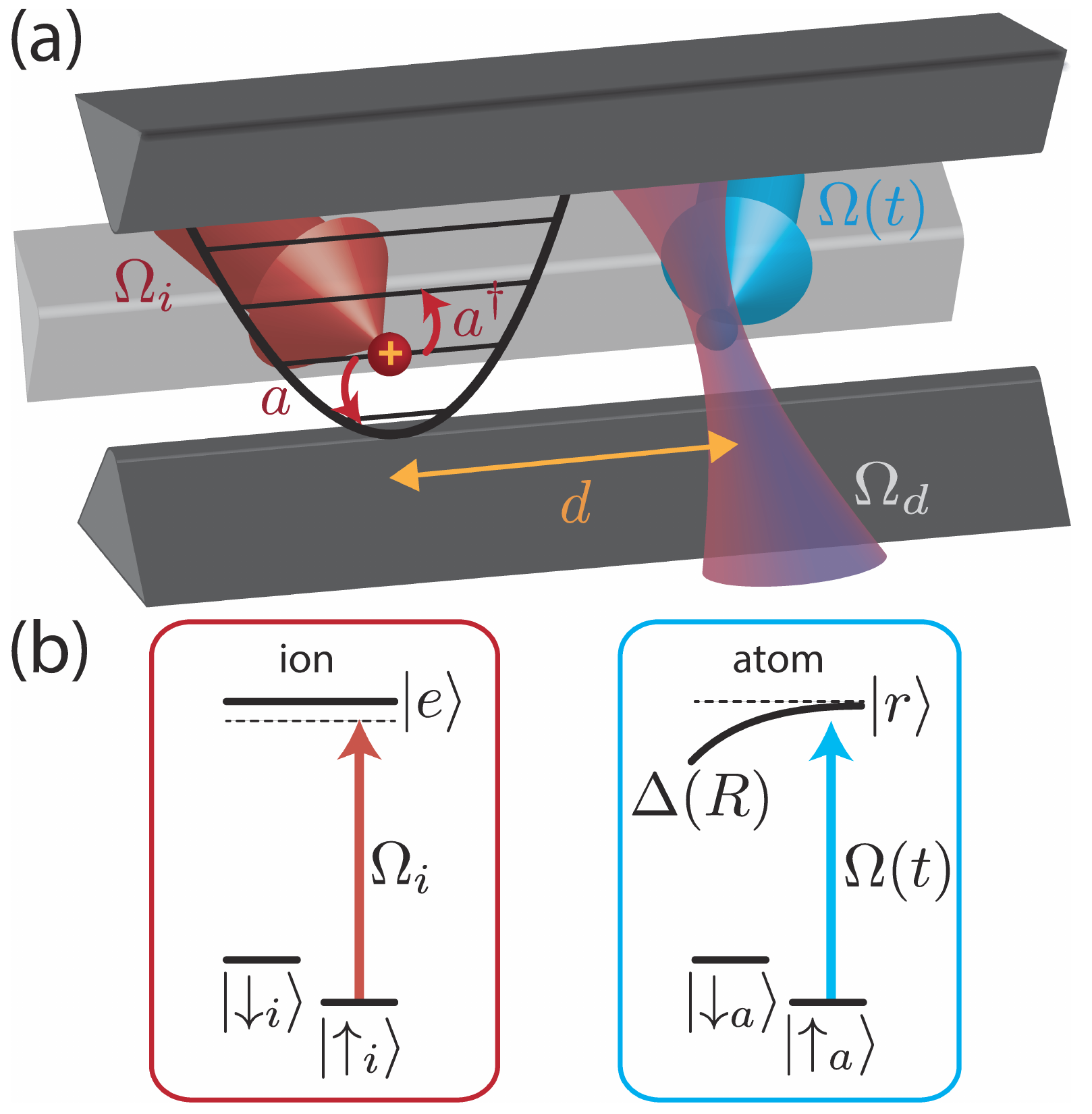}
\caption{ (a) We consider an ion (red ball) trapped in a Paul trap (grey electrodes) experiencing a harmonic confinement with trapping frequency $\omega_i$ and lowering (raising) operators $a$ ($a^\dag$).  A distance $d$ away, an atom (blue ball) is optically trapped with Rabi frequency $\Omega_d$ and coupled to a Rydberg state with a time-dependent laser (blue arrow) of Rabi frequency $\Omega(t)$. (b) Internal atomic (right) and ionic (left) level scheme. The Rydberg state experiences a position dependent Stark shift $\Delta (R)$, with $R$ the distace between the atom and the ion, due to the electric field of the ion. The resulting force may be used to entangle the atom and the ion when coupling to the Rydberg state depending on the internal state of the atom as described in the text.}
\label{fig_coordinates} 
\end{figure}

Here, we propose another route towards interacting atom-ion quantum systems and provide a solution to the micromotion-induced limitations when coupling these systems. In particular, we propose to couple atoms to low-lying Rydberg states~\cite{Henkel:2010,Pupillo:2010,Cinti:2010,Maucher:2011,Henkel:2011,Mobius:2013,Mattioli:2013,Macri:2013,Cinti:2014,Balewski:2014} such that their polarizability is significantly increased. This in turn increases the range over which the atom-ion interaction is effective to the micrometer range. Optical or magnetic potentials that bring the atoms within this distance of the ion, but still prevent the atoms from sampling the micromotion, can be straight-forwardly generated. Furthermore, the interaction can be made state-dependent by tuning laser parameters and allows for dynamical control as well. This last point is of importance to bridge the energy differences between trapped atomic and ionic systems commonly encountered in laboratory settings, by modulating the interaction close to the ionic motional resonance as we describe below. 

The interactions between the Rydberg-coupled atom and the ion can be used to entangle these particles or to mediate spin-spin interactions between atoms in close analogy to much employed quantum gates in trapped ions. The setup we have in mind is illustrated in Fig.~\ref{fig_coordinates}:  a single ion (red -- left) is trapped at the potential minimum of a Paul trap undergoing a harmonic oscillator motion with trap frequency $\omega_i$. In its close vicinity an atom (blue -- right) is optically trapped, experiencing both, the potential due to the Paul trap and the optical light field. A laser weakly couples one of the hyperfine ground states of the atom $\mid\uparrow\rangle_a$ to a Rydberg level, but not the second one $\mid\downarrow\rangle_a$. This can be done by, e.g., laser polarization or a large frequency difference between the states. By modulating the intensity of the Rydberg laser -- and thereby the atom-ion coupling -- at a frequency $\omega_{v}=\omega_i+\delta$ the ionic motion gets excited when the atom is in the state $\mid\uparrow\rangle_a$, but not when it is in the state $\mid\downarrow\rangle_a$. We will show that the effective interaction (to lowest order in the atom-ion separation and within the rotating wave approximation) can be described by:
\beq
\hat H_{I}\propto \left(\hat a^{\dag}e^{i\delta t}+\hat a e^{-i\delta t}\right) \mid \uparrow\rangle_a \langle \uparrow \mid,
\eeq
\noindent with $a^{\dag}$ and $a$ the creation and annihilation operators of the ionic motion. This effective Hamiltonian entangles the motional state of the ion with the internal state of the atom. An additional laser field that generates a spin-motion interaction of the form
\beq
\hat H_{S-M}\propto \left(\hat a^{\dag}e^{i\delta t}+\hat a e^{-i\delta t}\right) \mid \uparrow\rangle_i \langle \uparrow \mid,
\eeq
\noindent can be used to map the ion motion onto internal states of the ion, generating effective (pseudo) spin-spin interactions between atoms and ions~\cite{Porras:2004}. Alternatively, the ion motion may serve as a bus to generate spin-spin interactions between atoms. The scheme closely resembles state-of-the-art trapped ion based quantum gates and retains some of the main benefits associated with them~\cite{Sorensen:1999,Sorensen:2000,Leibfried:2003a,Roos:2008}. In particular, no ground state cooling of the atom or ion is required~\cite{Sorensen:1999,Sorensen:2000,Kirchmair:2009} and the gate and spin-spin interactions are to a large degree immune to micromotion~\cite{Landa:2014,Shen:2014}. In analogy to the trapped ion case, these benefits are a direct consequence of the (near) linearity of the atom-ion interactions at large distances and break down for small atom-ion distances.

The paper is organised as follows: In Sec.~\ref{Sec_Rydion} we derive the form of the Rydberg atom-ion adiabatic interaction potential and we demonstrate that significant interactions can be obtained over distances of a $\mu$m instead of the 100~nm range of the ground state-atom ion interaction. In Sec.~\ref{sec:spin-spin}  we show how a single trapped ion can be entangled with an atomic qubit by modulating the Rydberg laser field. We study the effects of imperfect ion and atom cooling and the coupling to the trapping fields of the ion and show that the interactions are resilient to these effects. Finally, in Sec.~\ref{sec:outlook} we draw conclusions and discuss the prospects for scaling up.

\section{Rydberg atom-ion interactions}
\label{Sec_Rydion}

In the following section we derive an effective potential for the interaction of an alkali Rydberg-atom with a singly charged ion for large atom-ion distances.  Our strategy to solve the problem 
will be the following: We start with a three-body model system comprised of two singly charged spin-less particles -- the ion and the core of the atom -- and an electron. 
Then, we will use the Born-Oppenheimer approximation and expand the ion-electron and ion-core interaction to second order in the relative core-electron coordinate to obtain the dominant 
charge-dipole and charge-quadrupole interaction terms. With such approximated interactions between the three particles, 
we diagonalise the resulting Hamiltonian in a truncated basis of Rydberg wavefunctions obtained by means of the Numerov method. We show that for the atom-ion separations and parameters 
considered in this work, second order perturbation theory suffices and we discuss the effects of the trapping fields. Finally, we consider a situation in which a ground state atom is weakly dressed 
with a Rydberg state and derive the corresponding adiabatic potential.

\subsection{The model Hamiltonian}

The dynamics of the system shown in Fig.~\ref{fig_coordinates} can be described by the Hamiltonian
\beq
\begin{aligned}
\label{eq:Hfull}
\hat{{H}}=&\hat{{H}}_a+\hat{{H}}_i+\hat{{H}}_{ia}+\hat{{H}}^{t}_a+\hat{{H}}_L.
\end{aligned}
\eeq
accounting for the dynamics of a free atom $\hat H_a$, the single trapped ion $\hat H_i$, the atom-ion interaction $\hat H_{ia}$, and the interaction of the atom with the fields of the Paul trap $\hat{{H}}^{t}_a$. The last term, $\hat{{H}}_L$,  describes the interaction of the atom with the Rydberg laser and the optical dipole trap. In the following we discuss each of these terms individually.

The first term of Eq.~\eqref{eq:Hfull} describes the dynamics of the free alkali atom. It possesses a single valence electron with the remaining electrons forming closed shells. For the description of such system one can employ an effective two-body approach in which the atom is modelled by a positively charged core of mass $m_c$ at position $\mathbf{r}_c$ and the single valence electron of mass $m_e$ at position $\mathbf{r}_e$. The Hamiltonian of this core-electron system is
\beq
\label{eq:Hatom}
\begin{aligned}
\hat{{H}}_a&=\frac{\pp_c^2}{2 m_c}+\frac{\pp_e^2}{2 m_e}+V_{Ryd}(\rr_e-\rr_c)-\frac{\pp_e^2\pp_e^2}{8m_e^3 c^2}+V_{SO}^{e-c}.
\end{aligned}
\eeq
It consists of the kinetic energies of the atomic core and electron, where $\hat{\mathbf{p}}_c$ and $\hat{\mathbf{p}}_e$ denote the momentum operators of the atomic core and electron, respectively. The interaction between the electron and the core can be modelled by $V_{Ryd}$~\cite{Marinescu:1994},  which depends on the (relative) positions of the electron $\mathbf{r}_e$ and core $\mathbf{r}_c$ and its angular momentum state. The last two terms take into account the relativistic correction to kinetic energy and the electron-core spin-orbit interaction
\beq
\begin{aligned}
\label{eq:SO1}
V_{SO}^{e-c}&=\frac{1}{2 m_e^2 c^2 }\, \bs\cdot\left[\nabla_{e}V_{Ryd}(\rr_e-\rr_c)\times\pp_e\right]
\end{aligned}
\eeq
giving rise to the fine-structure of electronic levels,
with $\bs$ the spin-1/2 operator of the electron and $\nabla_{e}$ the gradient with respect to the position of the electron.

The second term of Eq.~\eqref{eq:Hfull} describes the dynamics of the ion of mass $m_i$ at position $\mathbf{r}_i$ in the Paul trap,
\beq
\label{eq:Hion}
\hat{{H}}_i=\frac{\pp_i^2}{2 m_i}+e\Phi_{PT}(\rr_i,t)+\hat{{H}}_i^{int}+\hat{{H}}_i^{laser},
\eeq
with $\Phi_{PT}(\rr_i,t)$ the electric potential of a standard Paul trap (quadrupole field), $e$ the elementary charge and $\hat{\mathbf{p}}_i$ the momentum operator of the ion. The parameters of the Paul trap are chosen such that stable trapping is ensured and the ionic motion can be decomposed in a slow secular motion with trapping frequencies $\omega_i^{x,y,z}$ and a fast micromotion at the trap drive frequency $\Omega_{\rm rf}$~\cite{Leibfried:2003}.  The effect of this micromotion on the atom-ion interaction will be discussed in Sec.~\ref{sec:mm}.
The terms $\hat{{H}}_i^{int}$ and $\hat{{H}}_i^{laser}$ account for the internal electronic structure of the ion and the ion-laser interaction. We can treat the external and internal degrees of the ion separately as the internal state of the ion does not couple to the trapping fields for the low lying states considered here. We note that, in writing Eqs.~\eqref{eq:Hatom} and \eqref{eq:Hion} we neglected any cross-couplings of the Rydberg laser fields on the ion and the ion-laser fields on the atom, as the atom and ion considered have very different electronic level structures. 

The third term of Eq.~\eqref{eq:Hfull} describes the atom-ion interaction, which can be split into charge-charge interactions of the form $V_C(\mathbf{x})=e^2/(4\pi\epsilon_0|\mathbf{x}|)$ between the ion-core and ion-electron,
\beq
\label{eq:Hai}
\hat{{H}}_{ia}=V_C(\rr_c-\rr_i)-V_C(\rr_e-\rr_i)+V_{SO}^{e-i},
\eeq
Additionally, the electric field of the ion generates a spin-orbit like interaction for the electron
\beq
\label{eq:SO2}
V_{SO}^{e-i}=-\frac{1}{2 m_e^2 c^2 }\, \bs\cdot\left[\nabla_eV_C(\rr_e-\rr_i)\times\pp_e\right]
\eeq
which modifies the fine-structure of the electronic levels given by Eq.~\eqref{eq:SO1}. Since in the present analysis the hyperfine splitting between Rydberg states is smaller than the interaction energies and the laser detunings considered in this work,  it can safely be neglected.

\subsection{Born-Oppenheimer approximation}\label{subsec:BO}
In this section we study the interaction between a Rydberg atom (consisting of a core and the highly excited single valence electron) and the ion. In order to simplify the following presentation we first neglect internal degrees of the ion and any laser interactions between them, i.e. $H_i^{laser}=H_i^{int} =0$. For the same reason, we set $H_a^t=0$ in this section. The effect of the Paul trap on the atom, i.e. $H_a^t$, will be discussed in Sec. ~\ref{sec:atompaul}, where we show that they are small for distinct atom-ion separations. In particular, we derive Born Oppenheimer (BO) energy surfaces, which -- in a secular approximation -- give rise to effective interaction potentials between the Rydberg atom and the ion. These potentials form the basis for the further analysis of Rydberg dressing in Sec.~\ref{sec:dressing}.

For the following derivation it is convenient to change the frame of reference to the atomic center-of-mass (COM) coordinate $\rr_a=(m_e\rr_e+m_c\rr_c)/M$ and the relative electron-core coordinate $\rr=\rr_e-\rr_c$ with $M$ the total mass. The corresponding COM and relative momentum operators are $\pp_a$ and $\pp$, respectively.
First we rearrange terms in the full Hamiltonian 
\beq\label{eq:BOFull}
\hat{H}=\hat{H}_a+\hat{H}_i+\hat{H}_{ia}\approx\hat{H}_i+\frac{\pp_a^2}{2 M} +\hat{H}_{BO}
\eeq

\begin{figure}[t!]
\centering
\includegraphics[width=8.5cm]{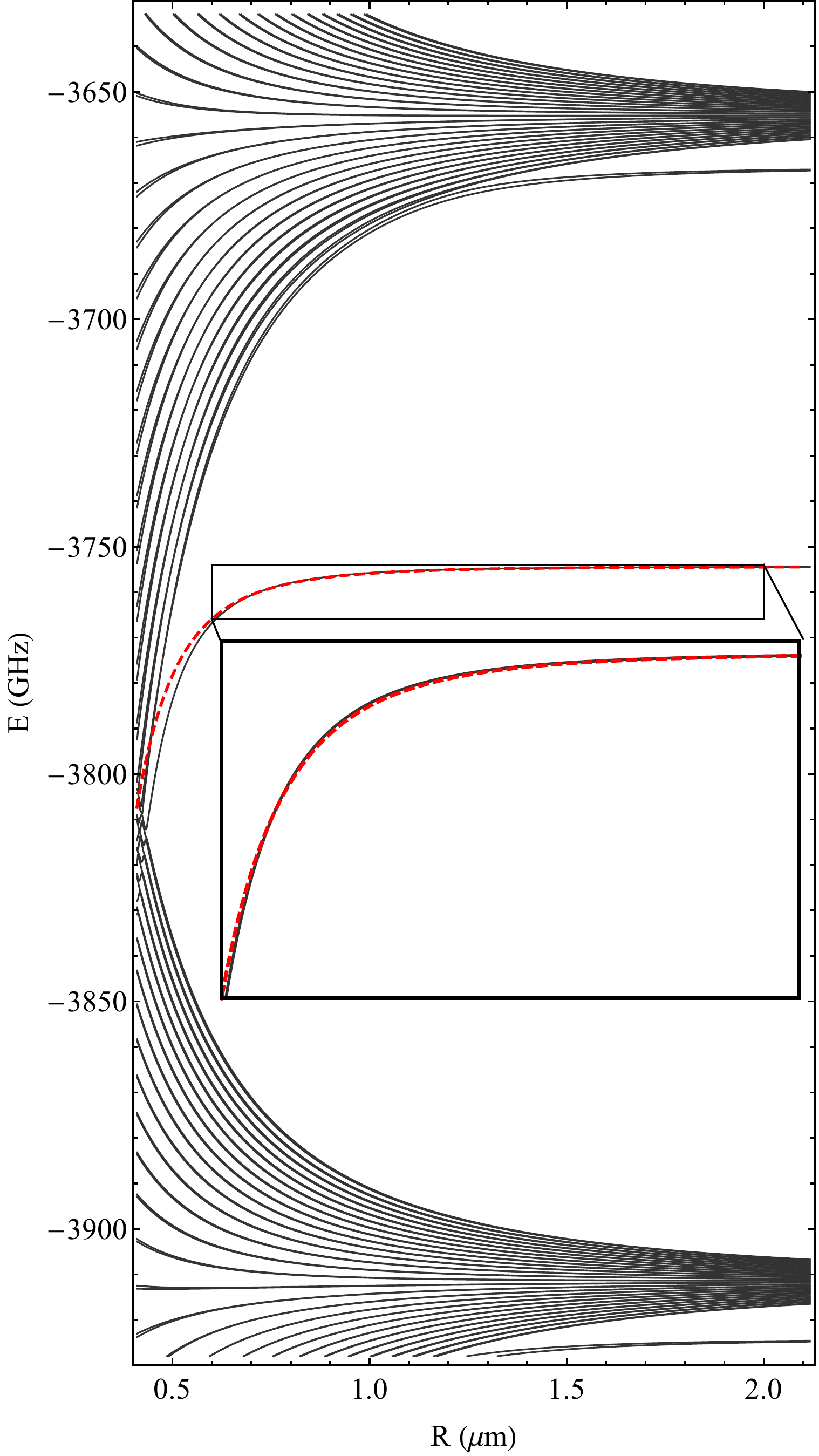}
\caption{ Eigenenergies $\epsilon_k$ of $\hat{H}_{BO}$ for $^6$Li interacting with an ion as a function of the ion-core distance $R$, that emanate from the $n=30$ and the $n=29$ manifolds based on our simulation without trapping fields.  The 30$S$ and 30$P$ energies lie separated at $-3754.4$ GHz and $-3666.7$ GHz. The dashed red line shows a $-C_4^{\mid 30S_{1/2}\rangle}/R^4$ potential shifted down by the $30S$ energy that is based on second order perturbation theory within the dipole approximation. Here $C_4^{\mid 30S_{1/2}\rangle}$ was taken from Refs.~\cite{Miffre:2005},~\cite{Kamenski:2014}. We see that the 30$S$ state remain well separated from the others down to distances of $\sim$~500~nm, whereas second order perturbation theory works well for distances $\geq 1 \mu$m. }
\label{fig:spaghetti} 
\end{figure}

\noindent We note that in writing Eq. (\ref{eq:BOFull}) we have neglected a component of the spin orbit terms as well as a contribution to the relativistic kinetic energy correction, which are small due to a prefactor of $m_e/M$ (see Appendix \ref{app:Rydint}). The BO Hamiltonian $\hat{H}_{BO}$ is defined as~\footnote{Here $\nabla V$ denotes the gradient of the potential function $V$, which results from first equating $\nabla_e V(\rr_e-\xx)=\nabla V(\rr_e-\xx)$  and transforming to center-of-mass and relative coordinates afterwards.}
\begin{widetext}
\beq
\begin{aligned}
\hat{{H}}_{BO}=&\frac{\pp^2}{2 \mu}+V_{Ryd}(\rr)-\frac{\pp^2\pp^2}{8m_e^3 c^2}+\frac{1}{2 m_e^2 c^2 }\, \bs\cdot\left[\nabla V_{Ryd}(\rr)\times\pp\right]\\
&-V_C\left(\RR+\frac{m_c}{M}\rr\right)
+V_C\left(\RR-\frac{m_e}{M}\rr\right)
-\frac{1}{2 m_e^2 c^2 }\, \bs\cdot\left[\nabla V_C\left(\RR+\frac{m_c}{M}\rr\right)\times\pp\right],\\
\label{eq:H_BO}
\end{aligned}
\eeq
\end{widetext}
where $\mu=m_e\,m_c/M$ is the reduced mass, and $\RR=\rr_a-\rr_i$ is the atom-ion separation. Treating $\rr_i$ and $\rr_a$ as parameters we can interpret $\hat H_{BO}$ as a family $\hat H_{BO}(\mathbf{r}_i,\rr_a)$ of operators on the Hilbert space of the relative coordinate $\rr$ only. The BO potentials $\epsilon_k(\mathbf{r}_i,\rr_a)$ are obtained as the eigenenergies of the BO Hamiltonian for fixed atom-ion distances and zero kinetic energies. We assume that the linewidths of the corresponding states are smaller than the energy separation between them, and that the relevant kinetic energies are small enough, such that Landau-Zener transitions between different BO surfaces can be neglected (secular approximation). In this case the resulting position dependent eigenvalues act as potentials in each state manifold. This yields effective Hamiltonians for the electron staying in the $k$-th energy level:
\beq
\label{eq:H_eff}
\begin{aligned}
\hat{{H}}_{\rm eff}^{(k)} =&\frac{\pp_i^2}{2 m_i}
+\frac{\pp_a^2}{2 M}
+\epsilon_k(\mathbf{r}_i,\rr_a)
+e\Phi_{PT}(\rr_i,t).
\end{aligned}
\eeq
Figure~\ref{fig:spaghetti} shows a typical example of the BO potentials around the Rydberg state $|30S_{1/2}\rangle$ of $^6$Li. The adiabatic eigenenergies are obtained by expanding the atom-ion interaction terms of the Hamiltonian $H_{BO}$ of Eq.~\eqref{eq:H_BO} up to second order in $|\rr|/R$ (taking into account the dominant charge-dipole and charge-quadrupole interactions) and diagonalizing it in the basis of unperturbed atomic wavefunctions (see Appendix~\ref{app:Rydint}). In particular, we see that the state $|30S_{1/2}\rangle$ remains well separated in energy  from the other states down 
to atom-ion distances of about 500~nm. For distances in the $\mu$m range, the full diagonalization is in excellent agreement with second order perturbation theory within the dipole approximation, which yields a potential of 
the form $-C_4^{|R\rangle}/R^4=-\alpha_{|R\rangle} \mathbf{E}_{ion}^2(\mathbf{R})/2 $ with $\mathbf{E}_{ion}(\mathbf{R})$ the electric field of the ion evaluated at the atom position, and $\alpha_{|R\rangle}$ the polarizability of the Rydberg state~\cite{Hahn:2000}. 
Since the polarizability scales with the principle quantum number to the power 7, $\alpha_{|R\rangle} \propto n^7$, it can be many orders of magnitude larger for the Rydberg state than for the ground state atom. For instance, for lithium in the $|30S_{1/2}\rangle$ state, $\alpha_{|R\rangle}$~=~3.5 $\times$10$^8$~$\alpha_{|2S_{1/2}\rangle}$~\cite{Miffre:2005,Kamenski:2014}. Note that the electron orbit in a Rydberg-atom is given by $r_n = n^2 a_0$ with $a_0$ being the Bohr radius. For $n=30$ 
we have $r_{30}\simeq 0.05\,\mu$m, which is indeed much smaller than the atom-ion separation ($\sim 1\,\mu$m) we are interested in.

\subsection{Atom-Paul trap interactions}
\label{sec:atompaul}
In this section we discuss the effect of the ionic trapping field from the Paul electrodes on the Rydberg atom. The interactions between the atom and the Paul trapping fields is described by
\beq
\hat{{H}}_a^{t}=e\Phi_{PT}(\rr_c,t)-e\Phi_{PT}(\rr_e,t)+V_{SO}^{e-t},
\eeq
where we included the spin-orbit potentials due to the trapping fields (with the superscript $t$ denoting `trap')
\beq
V_{SO}^{e-t}=\frac{e}{2 m_e^2 c^2 }\, \bs\cdot\left[\mathbf{E}_{PT}(\rr_e,t)\times\pp_e\right].
\eeq
Here, $\mathbf{E}_{PT}(\rr_e,t)=-\nabla_e \Phi_{PT}(\rr_e,t)$ denotes the electric field of the Paul trap at the electron position.
A linear Paul trap is generated by the electric fields $\mathbf{E}_{PT}(\mathbf{r},t)=\mathbf{E}_{\rm s}(\mathbf{r})+\mathbf{E}_{\rm rf}(\mathbf{r},t)$ with

\begin{eqnarray}
\mathbf{E}_{\rm s}(x,y,z)&=&\frac{m_i\omega_i^2}{e}\left(\frac{x}{2},\frac{y}{2},-z\right)\label{eqS},\\
\mathbf{E}_{\rm rf}(x,y,z,t)&=&\frac{m_i\Omega_{\rm rf}^2q}{2e}\cos \Omega_{\rm rf} t\left(x,-y,0\right)\label{eqRF},
\end{eqnarray}
and $q$ the stability parameter for an ion of mass $m_i$ and $\Omega_{\rm rf}$ the trap drive frequency. 
 
As it is clear from Eq.~(\ref{eqRF}), there is no oscillating field in the $z$-direction and the confinement along this axis is supplied by the static field, which generates a harmonic trap with a trap frequency $\omega_i$. The distance at which the trapping fields cancel the field of the ion in the $z$-direction  - assuming the ion is bound to the center of the trap $\mathbf{r}_i=0$ - is given by $\ell_z=\left[e^2/(4\pi\epsilon_0 m_i \omega_i^2)\right]^{1/3}$. For $^{171}$Yb$^+$ with $\omega_i=2\pi$~250~kHz, we have $\ell_z=$~6.9~$\mu$m.  For $z\ll \ell_z$, we can neglect the Stark shift of the static trapping field on the Rydberg level, whereas for $z\gg \ell_z$, the Stark shift of the trapping field dominates. For the transverse directions, the static field adds to the ionic field, and 
no cancelation occurs. The lowest value of the combined field occurs at $\ell_{\perp} = 2^{2/3}\ell_z$. 

The oscillating field $\mathbf{E}_{\rm rf}(x,y,z,t)$ supplies the confinement in the transverse direction. The motion of the ion in the transverse direction is given by a slow secular motion of frequency $\omega_i^{(\perp)}\approx \frac{\Omega_{\rm rf}}{2}\sqrt{a+q^2/2}$ 
with $a=-2\omega_i^2/\Omega_{\rm rf}^2$, and a fast micromotion of frequency $\Omega_{\rm rf}$. The amplitude of $\mathbf{E}_{\rm rf}(x,y,z,t)$ is typically a factor $\sim$~10-100 larger than the static field. Therefore, the effect of the ion trapping 
fields can only be neglected when considering atoms that are trapped close to the radio frequency null line $x,y\sim 0$. For the numbers considered in this work ($\Omega_{\rm rf}=2\pi$~2.5~MHz and $q=0.28$), the oscillating field at maximal amplitude starts to dominate over the ion field at 2.9~$\mu$m. From Fig.~\ref{fig:spaghetti} we see that this is within the range where second order perturbation theory can be used. Furthermore, since the energy gaps between the Rydberg states lie in the 100~GHz range, the MHz trapping field cannot drive transitions between the Rydberg states allowing us to treat the effect quasi-statically.

\subsection{Atom laser interactions: Dressed atoms}
\label{sec:dressing}
We end this section by considering the interaction with laser fields. The situation we have in mind is one where a ground state atom is weakly dressed by two laser fields. One of them, $\mathbf{E}_{\rm dip}(\rr_a,t)$, is a tightly focussed laser that creates the atomic trapping potential, whereas the other, $\mathbf{E}_{\rm dress}(\rr_a,t)$, couples the atom off-resonantly to a Rydberg state. Within the dipole approximation, the Hamiltonian is given by:

\beq
\label{eq:trap-atom}
\hat{{H}}_L=e\rr\cdot\left(\mathbf{E}_{\rm dress}(\rr_a,t)+\mathbf{E}_{\rm dip}(\rr_a,t)\right),
\eeq

We assume that each of the fields is tuned close to a single transition with low enough coupling strength to allow us to neglect all other transitions. In particular, the dipole laser is tuned close to a dipole allowed transition $|g\rangle \leftrightarrow |e\rangle$, with $|e\rangle = |2P\rangle$, whereas the Rydberg dressing laser is tuned close to the transition $|g\rangle \leftrightarrow |R\rangle$, where $|g \rangle$, $|e \rangle$ and $|R\rangle$ denote ground, excited and Rydberg state, respectively. Note that in practice, the Rydberg laser may be comprised of two light fields, to couple the $S$ ground state to some Rydberg state $nS$ via a $P$ state. We assume that the laser fields have (effective) Rabi frequencies of $\Omega_d (\rr_a)\propto E^0_{\rm dip}$ and $\Omega\propto E^0_{\rm dress}$, where $E^0_{\rm dress}$ and $E^0_{\rm dip}$ denote the electric field amplitudes of the two laser fields, and that they are detuned by $\Delta_d$ and $\Delta_0$ from the states $|e\rangle$ and $|R\rangle =  |nS_{1/2}\rangle$, respectively. A closeby ion causes a Stark shift in the Rydberg state such that the total detuning of the Rydberg state is given by $\hbar\Delta(R)=\Delta_0+\alpha_{|R\rangle}\vert \mathbf{E}_{ion}(|\mathbf{R}|)\vert^2/2$~\cite{Hahn:2000}. Here, we neglected interactions of higher order than the dominant charge-induced dipole interactions and spin-orbit interactions, as well as possible position dependence in $\Omega$. Within the rotating wave approximation, we can write the three-level interaction Hamiltonian in the $|g\rangle$, $|e\rangle$, $|R\rangle$ basis as:

\begin{equation}
\label{eq:H3level}
H_{3-level}=\left( \begin{array}{ccc}
0 & \hbar\Omega_d(\rr_a) & \hbar\Omega \\
\hbar\Omega_d (\rr_a) & -\hbar\Delta_d & 0\\
\hbar\Omega & 0 &-\hbar\Delta_0 -\frac{C_4^{|R\rangle}}{R^4} \end{array} \right).
\end{equation}
Here, we neglected the atom-ion interaction for the states $|g\rangle$ and $|e\rangle$, which is justified for $R\gg R^*$ with $R^*= (2\mu_{ai} C_4^{|g\rangle}/\hbar^2)^{1/2}$ the typical length scale of the ground state atom-ion interaction, with $\mu_{ai}$ the reduced atom-ion mass. For typical atom-ion combinations $R^*$ lies in the 100~nm range~\cite{Idziaszek:2007}. Assuming $|\Delta_0| \gg |\Omega|$, $\Delta_0>0$, i.e. blue detuning as well as $|\Delta_d| \gg |\Omega_d (\rr_a)|$ and $\Delta_d<0$, the Hamiltonian can be diagonalised to second order in $\Omega$ and $\Omega_d (\rr_a)$ to obtain the adiabatic potential $V_{ad}=V_{\rm dip}(\rr_a)+V(R)$. Here, $V_{\rm dip}(\rr_a)=\hbar|\Omega_d (\rr_a)|^2/\Delta_d$ represents the dipole trap, which we assume traps the atom harmonically with trap frequencies $\omega_a^{x,y,z}$, and

\beq\label{eqVad}
V(R)=-\frac{A R_w^4}{ R^4+ R_w^4}
\eeq 
\noindent denotes the dressed atom-ion potential.  The depth of this potential is given by $A=\hbar\Omega^2/\Delta_0$ and its width by $R_w=(C_4^{|R\rangle}/\hbar\Delta_0)^{1/4}$. We note that Eq.~(\ref{eqVad}) takes a similar form as the case for the atom-atom dressed Rydberg potential, see e.g.~\cite{Macri:2013}, but retains a $R^{-4}$ character instead of the $R^{-6}$ Van der Waals case of the atom-atom interaction, and it is always attractive for $|nS_{1/2}\rangle$ states. Note that the potential is also of lower order, scaling as $\Omega^2/\Delta_0$ instead of $\Omega^4/\Delta_0^3$, because for the ion-atom case only a single particle needs to be dressed. This relaxes restraints on the required laser power. For red detunings, the potential is also attractive, but an avoided crossing occurs at $R=R_w$, such that resonant Rydberg excitation may result. 

Taking the ion trapping fields into account within second order perturbation theory and retaining only the charge-dipole terms, the adiabatic potential~(\ref{eqVad}) is changed to:

\begin{equation} 
\tilde{V}(\mathbf{r}_i,\mathbf{r}_a)=\frac{\hbar\Omega^2}{\Delta_0+\frac{\alpha_{|R\rangle}}{2\hbar}\left\vert\mathbf{E}_{ion}(|\mathbf{r}_i-\mathbf{r}_a|)+\mathbf{E}_{PT}(\mathbf{r}_a,t)\right\vert^2}.
\end{equation}
Since the Stark shift due to the electric fields always increases, the frequency offset from resonance for blue detuning, $\Delta_0 \gg |\Omega|$, assures adiabaticity.

\begin{figure}[hbt]
\centering
\includegraphics[width=8cm]{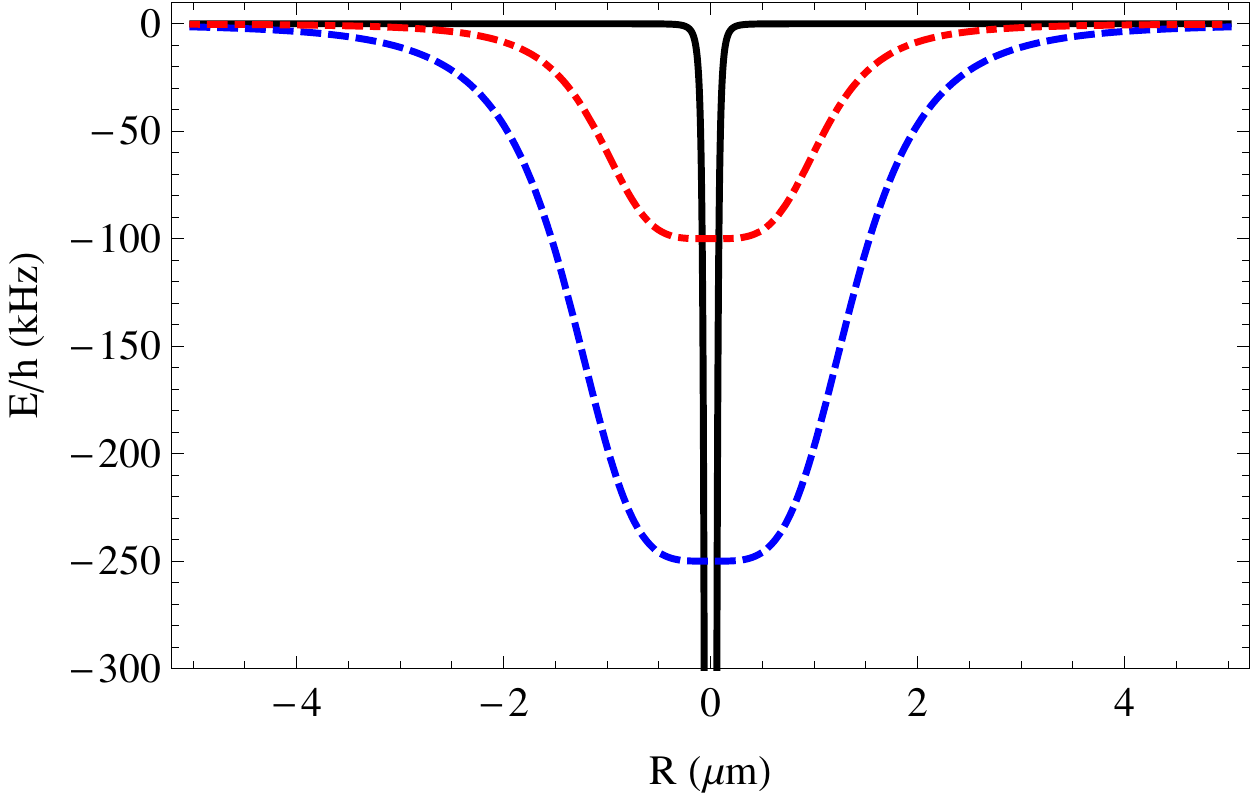}
\caption{Adiabatic potentials for a ground state atom and an ion (solid black), for a dressed atom with $\Omega=2\pi$~10~MHz and $\Delta=2\pi$~1~GHz (red dash-dotted) and $2\pi$~0.4~GHz (blue dashed) assuming coupling to the $\mid 30S_{1/2}\rangle$ state of lithium.}
\label{fig_dress} 
\end{figure}

To conclude this discussion, let us consider a lithium atom with $n=30$, $\Omega = 2\pi$~10~MHz, $\Delta_0=2\pi$~1~GHz, for which we have $A/h=$~100~kHz and $R_w=1$~$\mu$m, such that $R_w \gg R^*$ (e.g., assuming an Yitterbium ion). For these numbers, the lifetime of the dressed atom is enhanced by a factor 10$^4$ as compared to the Rydberg state, putting coherent experiments on the 100 ms timescale within reach. In Fig.~\ref{fig_dress} we show the resulting adiabatic potential. 
As it is shown, the adiabatic potential of a Rydberg dressed atom discussed above (red dash-dotted and blue dashed lines) has a much longer-ranged character than the corresponding ground 
state atom-ion interaction (solid black line). It also shows that at intermediate distances, 1-2 $\mu$m, the interaction is to a good approximation linear with respect to the atom-ion separation. This will be a 
crucial element for the implementation of quantum gates and the impact of the ionic micromotion on the atom, as we shall discuss in the next sections.

\section{Atom-ion spin-spin interactions}
\label{sec:spin-spin}

In the following, we show how an atomic qubit can be entangled with a single trapped ion. The atomic pseudospins are encoded in two long lived hyperfine states $\mid\downarrow\rangle_a$ and $\mid\uparrow\rangle_a$ as shown in Fig.~\ref{fig_coordinates}(b).
Choosing proper hyperfine states and laser polarizations (or by employing frequency differences 
due to hyperfine or fine structure splittings) one can achieve that the laser couples only the state $\mid\uparrow\rangle_a$ to a Rydberg level, leaving the state $\mid\downarrow\rangle_a$ unaffected.

We assume the atom to be trapped at some distance $d \gg R^*$, but $d\ll \ell_z$, away from the ion, along the $z$-axis of the ion trap, such that we can neglect for now the effect of the trapping potential for the ion on the atom and also the ground 
state interaction between the atom and ion, as justified in Sec.~\ref{sec:atompaul}. Then, we couple the $\mid\uparrow\rangle_a$ state of the atom to a Rydberg state with Rabi frequency $\Omega$ and detuning $\Delta_0$, such that their interaction Hamiltonian is given 
by $\hat H_{ai}=V(\hat{z}_i-\hat{z}_a+d) \mid\uparrow\rangle_a  \langle \uparrow \mid$ with $V$ given by Eq.~(\ref{eqVad}) and $\hat z_i$ and $\hat z_a$ the ion and atom position with respect to their equilibrium position. We now expand this potential around the ion and atom equilibrium positions $\bar{z}_i=0$ and $\bar{z}_a=0$: 
$V\approx V(d)+F_0z_i-F_0z_a+....$, with

\begin{eqnarray}
F_0&=&\left.\frac{dV}{dz_i}\right\vert_{z=\bar{z}}=-\left.\frac{dV}{dz_a}\right\vert_{z=\bar{z}},
\end{eqnarray}

\noindent where we used $\bar{z}\equiv(\bar{z}_i,\bar{z}_a)$ as a short hand notation for the two equilibrium positions. The force between the atom and ion reaches its highest value of $F_0=1.065 A/R_w$ for $d=0.88 R_w$, whereas the second order terms vanish at this point. 
Introducing the normal creation and annihilation operators for the atom ($\hat b ^{\dag}$ and $\hat b$) and ion  ($\hat a^{\dag}$ and $\hat a$) in their trap, we can write the full Hamiltonian as: $\hat H = \hat H_{trap} + \hat H_{ai}$ with 

\begin{eqnarray}
\hat H_{trap}&=&\hbar\omega_i\hat a^{\dag}\hat a+\hbar \omega_a \hat b^{\dag} \hat b\label{eq_H0},\\
\hat H_{ai}&\approx&\left[V(d) + F_0\ell_i(\hat a^{\dag}+\hat a) - F_0\ell_a(\hat b^{\dag}+\hat b)\right] \mid\uparrow\rangle_a  \langle \uparrow \mid.\nonumber
\end{eqnarray}
\noindent Here, $\ell_j=\sqrt{\hbar/(2m_j\omega_j)}$ for $j=i,a$, and $\mid\uparrow\rangle_a  \langle \uparrow \mid=(\hat\sigma_z^a+\mathbb{1})/2$, where $\hat\sigma_z^a$ denotes the Pauli matrix for the atom and $\mathbb{1}$ is 
the identity matrix. In order to induce large ion motion we modulate the force between the atom and ion close to the ionic trapfrequency. Therefore, we introduce time dependence in $A\rightarrow A(t)= A_0(1-\cos \omega_v t)/2$, by amplitude modulating the Rydberg laser, e.g. using an acousto-optical modulator~\cite{Kirchmair:2009}.  As long as $\omega_v \ll \Delta_0$, the minimal detuning of the laser, no resonant Rydberg excitation can occur and we can treat the modulation of the dressed potential quasi-statically. In order to obtain
the slowly changing dynamics we go into an interaction picture with respect to $\hat H_{trap}+V_{static}(d)\mid\uparrow\rangle_a  \langle \uparrow \mid/2$, with $V_{static}(d)$ denoting the static part of $V(z)$. Now, by defining $\delta = \omega_v -\omega_i$ and assuming 
$\hbar|\omega_v-\omega_a| \ll F_0\ell_a, F_0\ell_i$ we can make a rotating wave approximation by neglecting terms rotating faster than $\delta$ to obtain: 

\beq\label{Eq_atom_spin}
\hat H_I=\frac{F_0\ell_i}{4}\left(\hat a^{\dag}e^{i\delta t}+\hat a e^{-i\delta t}\right)\mid\uparrow\rangle_a  \langle \uparrow \mid.
\eeq

\noindent Here, we also assumed $\hbar\omega_v \gg V(d)/2$ such that fast oscillating position-independent Stark shifts in the atom average out. 

The Hamiltonian~(\ref{Eq_atom_spin}) entangles the motion of the ion with the internal state of the atom. In particular, for $\delta=0$, and starting from the state $\ket{\psi_{in}}=|0\rangle_{mi} (\mid\downarrow\rangle_a+\mid\uparrow\rangle_a)$, where $|0\rangle_{mi}$ denotes the ground state of ion motion, the Hamiltonian generates cat-like states of the form $\ket{\psi_{out}}=|0\rangle_{mi}|\downarrow\rangle_a+|\beta\rangle_{mi}\mid\uparrow\rangle_a$ after a time $t$ with $\mid\beta\rangle_{mi}$ denoting a coherent state of amplitude $\beta=F_0\ell_it/(4\hbar)$. 

Adding a bichromatic laser field that interacts with the internal states of the ion~\cite{Sorensen:1999,Sorensen:2000,Leibfried:2003a,Roos:2008} such that
\beq\label{Eq_ion_spin}
\hat H_{S-M}=\frac{\eta\hbar\Omega_{S-M}}{2}\left(\hat a^{\dag}e^{i\delta t}+\hat a e^{-i\delta t}\right)\mid\uparrow\rangle_i  \langle \uparrow \mid
\eeq
\noindent with the laser Lamb-Dicke parameter $\eta=\delta k\ell_i\ll 1$, $\delta k$ the wavenumber difference between the two components of the bichromatic laser field and $\Omega_{S-M}\propto \Omega_i^2$ its (effective) Rabi frequency, allows us to map the entanglement onto the internal state of the ion. This is most easily seen when we set $\eta\hbar\Omega_{S-M}/2=-F_0\ell_i/4$ in the total interaction Hamiltonian $\hat H_I+\hat H_{S-M}$. In this case no motion is excited in the ion when the spin of both atom and ion are down (since the operators $\mid\uparrow\rangle_a  \langle \uparrow \mid$ and $\mid\uparrow\rangle_i  \langle \uparrow \mid$ evaluate to zero for such states). When both particles are in the spin up state, also no motion is excited since the two forces cancel. Only when the particles have opposite spins is the ion motion excited and thereby the energy changed. This results in an interaction that is similar to the one usually encountered in M{\o}lmer-S{\o}rensen gates or phase gates in ions~\cite{Molmer:1999,Leibfried:2003a}. After a time $\tau=2\pi/\delta$ this accumulates in an effective interaction that is locally equivalent to $\hat H_{zz}=J\hat \sigma_z^i \hat \sigma_z^a/2$ with $J=F_0^2\ell_i^2/(32\hbar\delta)$ and the ionic motion returns to the initial orbit. Setting $J\tau/\hbar=\pi/4$ corresponds to a geometric phase quantum gate~\cite{Sorensen:1999,Leibfried:2003a}. The coupling of Eq.~(\ref{Eq_ion_spin}) can be obtained by two counter propagating beams with frequency difference $\omega_v$ that are far detuned from an excited state. By proper choice of polarization and states, this geometry can implement differential Stark shifts into the spin states of the ion of the form~(\ref{Eq_ion_spin})~\cite{Sorensen:1999,Leibfried:2003a}.

\begin{figure}
\centering
\includegraphics[width=8.5cm]{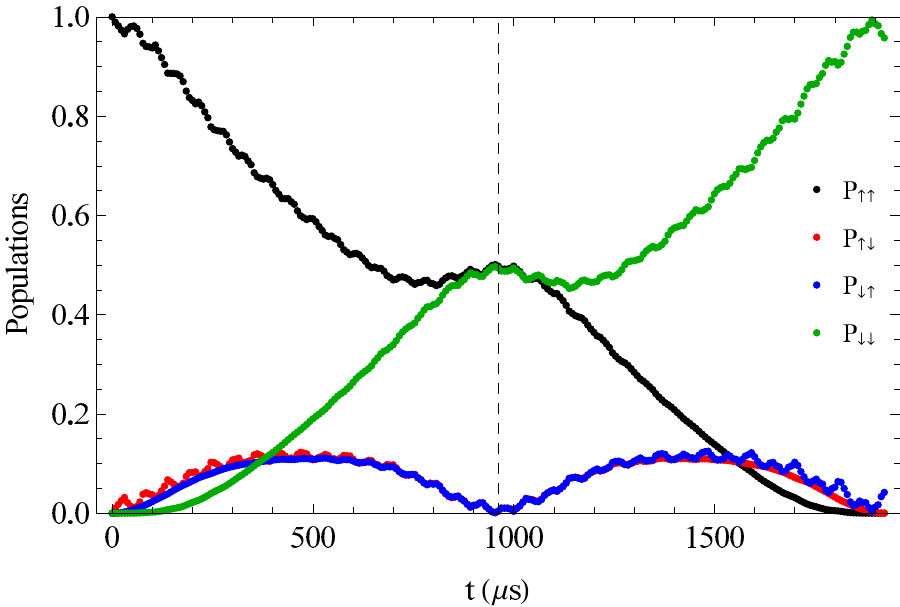}
\caption{Dynamics of the populations $P_{\uparrow\uparrow},..$ during the gate for the input state $\vert \psi_i \rangle=(|\uparrow\rangle_a+|\downarrow\rangle_a)(|\uparrow\rangle_i+|\downarrow\rangle_i)/2$ and after performing an additional $\pi/2$-pulse (see text). The dashed line indicates the time at which the gate is finished. The small oscillations are due to the term $V(d)|\uparrow\rangle_a  \langle \uparrow |/2$ that was neglected in the rotating wave approximation in deriving Eq.~(\ref{Eq_atom_spin}). For the present parameters we have that $V(d)/(2\hbar\omega_v) = 0.31$, but notwithstanding the fidelity of the gate is not seriously affected. }
\label{fig_dynamics} 
\end{figure}

We can also map the atom-ion entanglement on a second atom. For this, we consider two atoms, each trapped on one side of the ion along the $z$-axis of the ion trap. The Rydberg laser will once again only induce motion in the ion 
when the atoms are in opposite states and the effective interaction is proportional to $J\hat\sigma_z^a\hat\sigma_z^a$. In principle even more atoms may be involved, but the relative strengths of the spin-spin interactions will depend on the 
positioning of the atoms with respect to the ion. We will study this interesting many-body scenario in a future work. 

As a particular example, we consider a $^7$Li atom interacting with a $^{171}$Yb$^+$ ion. We set the trap frequency of the ion to $\omega_i=2\pi$~250~kHz and the trap frequency of the atom to $\omega_a=2\pi$~205~kHz. Using  $n=30$, $\Omega = 2\pi$~10.02~MHz, 
$\Delta_0=2\pi$~0.4~GHz, we have $A_0/h=$~250~kHz and $R_w=1.4$~$\mu$m. Further, for the ion laser driving field we use $\eta\Omega_{S-M}=2\pi$~1.045~kHz and $\delta=2\pi$~1.040~kHz, and we set the distance between the atom and ion trap 
to $d=0.88~R_w=1.23$~$\mu$m to optimise the coupling. To check whether the Rydberg state is still in the perturbative regime under these conditions [i.e. such that Eq.~(\ref{eqVad}) holds], we have numerically obtained the eigenstates of the lithium Rydberg states over the relevant distance by taking into account charge-dipole and charge-quadrupole interactions between the atom and ion, as we discussed in Sec.~\ref{Sec_Rydion}. We note that Rydberg excitation of individually trapped atoms with accurate positioning has been reported in a number of works~\cite{Urban:2009,Gaetan:2009,Saffman:2009}, such that the considered setup seems to be within experimental reach.

For the chosen parameters, the static trapping field of the ion can be neglected, but we have taken terms rotating faster than $\delta$ in the total Hamiltonian into account as well as terms up to third order in $\hat{z}_i$ and $\hat{z}_a$. 
More precisely, the simulated gate dynamics corresponds to the following Hamiltonian:

\begin{align}
\hat H_{g} = \hat H_{trap} + \frac{V^{(3)}(\hat{z}_i,\hat{z}_a)}{2}(1+\cos \omega_v t)\mid\uparrow\rangle_a  \langle \uparrow \mid+\hat{H}_{S-M}
\end{align}
where $V^{(3)}(\hat{z}_i,\hat{z}_a)$ denotes the Taylor expansion around the equilibrium positions up to third order of $V(\hat{z}_i,\hat{z}_a)$.
For simplicity, we have for now neglected the motion in the transverse direction and thereby the time-dependent electric fields. We consider the product input states $\vert\psi_i^{\pm \pm}\rangle=(\mid\uparrow\rangle_a\pm\mid\downarrow\rangle_a)(\mid\uparrow\rangle_i\pm\mid\downarrow\rangle_i)/2$, which can be prepared by simple radio-frequency pulses, and assume the motional ground states for the atomic and ionic oscillators. The case of $\vert\psi_i^{++}\rangle$ is shown in Fig.~\ref{fig_dynamics}. The motion of the ion returns to its initial orbit after $\tau_g = 2\pi/\delta=$~962.5~$\mu$s and the electronic state of the atom-ion system is found to be locally equivalent to the entangled state $\vert\Phi^+\rangle=(\mid\uparrow\rangle_a\mid\uparrow\rangle_i+i\mid\downarrow\rangle_a\mid\downarrow\rangle_i)/\sqrt{2}$ for the input state $\vert\psi_i^{++}\rangle$. This can be checked by performing the local unitary ($\pi/2$-pulse) $\hat{U}={\rm exp}(-i\pi (\hat\sigma_y^a+\hat\sigma_y^i)/4)$ to the state after the gate, leading to a fidelity of $F=0.997$. 
The fidelity is simply defined as the modulus square of the scalar product between the time evolved state $|\psi_{out}\rangle$ and the goal state $\vert0_i,0_a\rangle\otimes\vert\Phi^+\rangle$.
Similarly, the input states $\vert\psi_i^{+-}\rangle$, $\vert\psi_i^{-+}\rangle$ and $\vert\psi_i^{--}\rangle$ map to the entangled states $(\mid\uparrow\rangle_a\mid\downarrow\rangle_i\pm i\mid\uparrow\rangle_a\mid\downarrow\rangle_i)/\sqrt{2}$ and $(\mid\uparrow\rangle_a\mid\uparrow\rangle_i-i\mid\downarrow\rangle_a\mid\downarrow\rangle_i)/\sqrt{2}$, respectively following the gate and the unitary $\hat{U}$, all with fidelities $F\geq0.992$. We attribute the deviation from unit fidelity to interactions beyond linear and rotating terms neglected in the rotating wave approximation, but we expect that further parameter tuning, for example, via optimal control, can improve the fidelity further. The present goal, however, is to demonstrate that the proposed scheme is in principle possible, as the attained fidelities prove.

\subsection{Gate on thermal states}

As with ionic quantum gates that are essentially described by the same equations to first order, no ground state cooling is required, although we need the Lamb-Dicke regime for both the ion-laser and the atom-ion interaction, namely $\eta \ll 1$ and $F_0\ell_a \ll \hbar \omega_a$, $F_0\ell_i \ll \hbar \omega_i$. To investigate this property, we calculate the gate dynamics for the thermal input state 

\begin{equation}
\label{eq_rho_therm}
\hat\rho_{th}=\sum_{n_i,n_a} P_{n_i} (\bar{n}_i)P_{n_a} (\bar{n}_a)|n_i,n_a\rangle \langle n_i,n_a|\otimes|\psi^{++}\rangle \langle \psi^{++}|
\end{equation}
\noindent with $P_n (\bar{n})=\frac{1}{1+\bar{n}}\left(\frac{\bar{n}}{\bar{n}+1}\right)^n$ and $\bar{n}$ the average phonon number. 

When we start with both the atom and the ion in a thermal motional state with average phonon number $\bar{n}_i=\bar{n}_a=0.25$, the fidelity of the resulting Bell state is found to be $F=0.992$, demonstrating that the gate indeed works for non-ground state cooled particles too. 
We note that in this case the fidelity is defined as: $F= \mathrm{Tr}\{\hat \rho_g\hat\rho_{out}\}$, where $\hat\rho_g$ represents the goal state (e.g., $\hat\rho_g=\sum_{n_i,n_a} P_{n_i} (\bar{n}_i)P_{n_a} (\bar{n}_a)|n_i,n_a\rangle \langle n_i,n_a| 
\otimes\vert\Phi^+\rangle\langle\Phi^+\vert$) and $\hat\rho_{out}$ the output state after the gate and unitary $\hat{U}$.
We attribute the fidelity loss to the higher order terms in $\hat{z}_i$ and $\hat{z}_a$, as the linear approximation in the atom-ion interaction works less well for higher lying Fock states.  For the simulation, we limited the summation range in Eq.~(\ref{eq_rho_therm}) to $n_a,n_i=\{0,..,3\}$ in a Hilbert space that spans 9~phonons for both the atom and ion and such that Tr$(\hat\rho_{th})=0.997$.

\subsection{Micromotion}
\label{sec:mm}

To investigate the role of micromotion during the motional excitation of the ion we consider the situation in which the atom is trapped some distance $d$ away from the ion in the {\it transverse} direction. To deal with the micromotion of the ion, we replace the simple harmonic oscillator term in Eq.~(\ref{eq_H0}) of the ion with $\hbar\omega_i \hat a^{\dag}\hat a\rightarrow \hat H_{mm}(t)=m_i\Omega_{\rm rf}^2q\hat{x}_i^2\cos \left(\Omega_{\rm rf}t\right)/4+\hat{p}_i^2/(2m_i)$~\cite{Cook:1985}, where 
$\hat{p}_i=\frac{\hbar}{2i\ell_i}(\hat a^{\dag}-\hat a)$ is the ion momentum, $\hat{x}_i=\ell_i (\hat a^{\dag}+\hat a)$ its transverse position, and the trap drive frequency is given by $\Omega_{\rm rf}$. Note that in this Hamiltonian we use as a basis the harmonic oscillator states of $\hbar \omega_i a^{\dag}a$ where we set $\omega_i = \Omega_{\rm rf}q/2^{3/2}$ and neglect the static trapping field ($a=0$), which typically is a factor 10-100 smaller than the time-dependent field in a Paul trap. Besides this, we only consider the $x$-direction to reduce Hilbert space size, i.e. the problem is again one-dimensional. Additionally, instead of using $V(\mathbf{r})$, we take $\tilde{V}(\hat{x}_i,\hat{x}_a)$ that includes the Stark shift due to the time dependent trapping field. Classical simulations show that expanding $\tilde{V}(\hat x_i,\hat x_a)$ to third order in $\hat x_i$ and $\hat x_a$ for $d=1$~$\mu$m approximates the ion and atom orbits to within $< 5\times 10^{-3}\ell_{i,a}\sim 5$~pm. In total then the Hamiltonian for which the dynamics is simulated is given by:

\begin{equation}
\label{eq:mmHtot}
\hat H_{tot}(t) = \hat H_1+ \frac{\tilde{V}^{(3)}(\hat{x}_i,\hat{x}_a)}{2}(1+\cos \omega_v t)\mid\uparrow\rangle_a  \langle \uparrow \mid+\hat H_{S-M}.
\end{equation}
\noindent Here, $\tilde{V}^{(3)}(\hat{x}_i,\hat{x}_a)$ denotes the Taylor expansion around the equilibrium positions up to third order of $\tilde{V}(\hat{x}_i,\hat{x}_a)$ and $\hat H_1= \hat H_{mm}(t)+\hbar\omega_a (\hat b^{\dag} \hat b+1/2)$.

We again consider $^{171}$Yb$^+$ and $^7$Li coupled to $n=30$ and use the parameters $\omega_a=2\pi$~200~kHz, $\Omega_{\rm rf}=2\pi$~2.5~MHz, $q=0.28$ and $\eta\Omega_{S-M}=2\pi$~1.06~kHz and the (approximate) ground states 
of motion. We use classical physics to obtain the real secular trap frequency of the ion~\cite{Leibfried:2003} which is given by $\omega_i^{(\perp)}=2\pi$~254.089~kHz and we set $\delta^{(\perp)}=\omega_v-\omega_i^{(\perp)}=2\pi$~1.064~kHz 
(see also appendix~\ref{app:mm}).

\begin{figure*}
\centering
\includegraphics[scale=0.5]{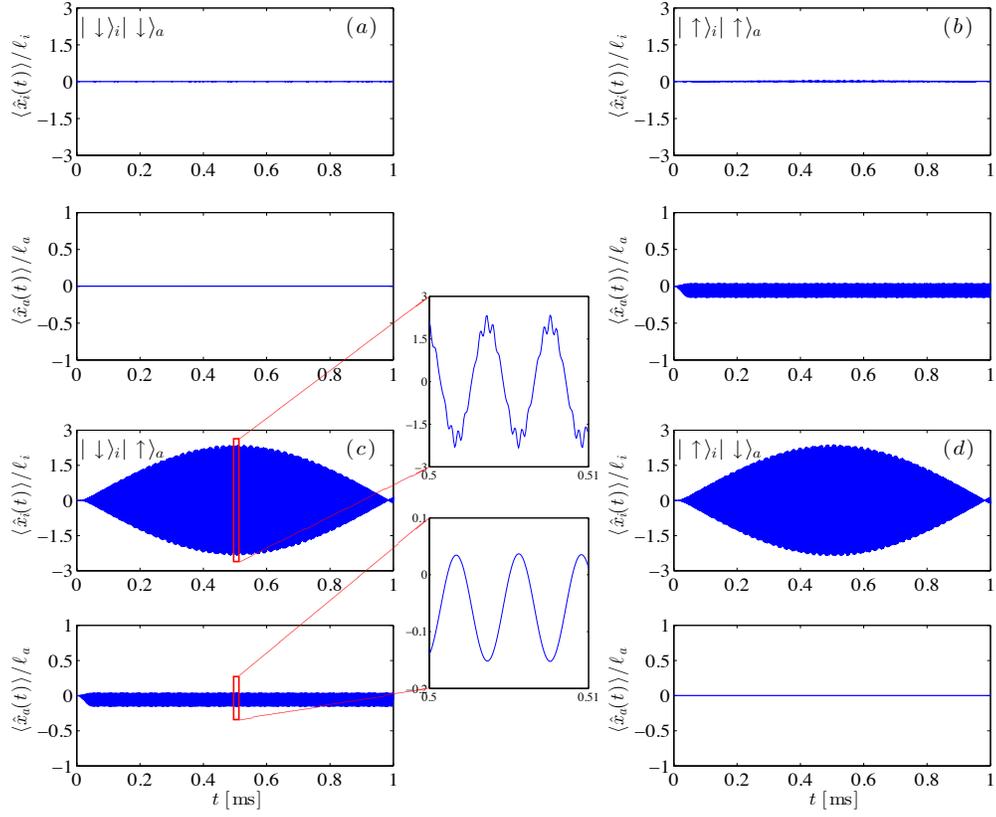}
\caption{ Expectation values $\langle\hat{x}_i\rangle$ and $\langle\hat{x}_a\rangle$ during the gate for each of the four possible spin input states. The insets also show zoom ins of the ion and atom motion, clearly showing the micromotion. Some residual atomic motion occurs for the states $|\uparrow \rangle_a$, where the atom gets pulled closer to the ion.}
\label{fig_gate_mm} 
\end{figure*}

For the Rydberg laser, we set $\Omega=2\pi\, 13.1$~MHz and $\Delta_0=2\pi$~0.8~GHz. To limit the induced motion in the atom, we switch on the Rydberg dressing in 50~$\mu$s (see Appendix~\ref{app:mm} for more details on the calculation). In 
Fig.~\ref{fig_gate_mm} we show the dynamics of the position expectation values for the atom and the ion for each of the possible spin states, demonstrating that the 
micromotion does not distort the motion of the particles during the gate significantly. As in the case without micromotion, the motion returns to its input state after the 
gate is finished, that is, in about $2\pi/\delta^{(\perp)}=940$~$\mu$s without additional energy exchange between the atom and ion, demonstrating the 
resilience of the scheme to micromotion.

The presented quantum gate closely resembles that of common ion gates, e.g.~\cite{Kirchmair:2009}. As with those gates, we can improve the fidelity by making sure the approximations made to obtain the gate dynamics - neglecting fast rotating terms and assuming the Lamb-Dicke regime - are well justified. This means for the atom that tight confinement needs to be reached. Furthermore, to reach a gate time that is much faster than the photon scattering rate $\Gamma_{ph}\sim(4\Delta_0^2/\Omega^2)\times \Gamma_{Ryd}$, strong laser fields are useful. In the present example, the lifetime of the bare Rydberg state lies in the 10-20~$\mu$s regime~\cite{Beterov:2009}, leading to lifetimes of $\sim$~100~ms~\cite{Balewski:2014} for the dressed case.

\section{Outlook}
\label{sec:outlook}

In conclusion, we have theoretically investigated the interaction between a single atom coupled to a Rydberg state and an ion trapped in a Paul trap. The large polarizability of the Rydberg state allows for strong interactions between the atom and the ion mediated over a $\mu$m. These interactions may be used to generate entanglement between the ion and the atom by state-dependent excitation of the ion motion. The resulting gate mechanism retains some of the benefits of the trapped ion platform, namely no need for ground state cooling, full dynamical control and near immunity to micromotion. The proposed scheme allows for interesting extensions such as dressing the atoms with a state of negative polarizability such that the atom-ion interaction becomes repulsive. This eliminates micromotion induced heating over an energy range set by the power available in the dressing field, even without tight atomic trapping potentials. Dressing with higher angular momentum states may also bring interesting functional forms of the atom-ion interaction potentials within reach, such as the non-dispersive charge-dipole $1/r^2$ and charge-quadrupole $1/r^3$ terms. Larger systems of interacting atoms and ions may also be considered in the future, in which not only atom-ion, but also ion-ion and atom-atom interactions become important. Finally, let us note that recent experiments with Rydberg {\it ions} in Paul traps demonstrate the feasibility of Rydberg excitations 
in the presence of the ion trapping field~\cite{Muller:2008,Schmidt-Kaler:2011,Feldker:2015}. Hence, given the similarities of our proposal and those recent experiments, the prospects of confining Rydberg-dressed atoms interacting with ions in a Paul trap are indeed very promising.\\

\acknowledgements
We gratefully acknowledge fruitful discussions with Peter Zoller. R.G. and T.S. acknowledge Krzysztof Jachymski, Thomas Feldker and Ferdinand Schmidt-Kaler for valuable comments. This work was supported by the EU via the ERC (Starting Grant 337638) and EQuaM (grant 323714) and the Netherlands Organisation for Scientific Research (NWO) (R.G.) and by the excellence cluster `The Hamburg Centre for Ultrafast Imaging - Structure, Dynamics and Control of Matter at the Atomic Scale' of the Deutsche Forschungsgemeinschaft (A.N.). A.W.G. acknowledges the SFB FoQuS  (FWF Project No. F4016-N23) and the ERA-NET CHIST-ERA (R-ION consortium) for support.

\onecolumngrid

\appendix
\section{Rydberg-ion interaction}
\label{app:Rydint}
\subsection{Born Oppenheimer approximation}
In this appendix we provide a detailed derivation of the Rydberg atom-ion effective Hamiltonian, whose derivation is based on the Born-Oppenheimer approximation.  
The idea is that for quantum systems comprised of particles that can be divided into two classes, light and heavy ones, one can approximately separate and solve first the dynamics of the light particles, namely by 
diagonalizing the Born-Oppenheimer Hamiltonian, which then, for each solution $\ket{\phi_k}$ of the light particle problem, yields an effective Hamiltonian $\hat{H}_{eff}^{(k)}$  for the heavy particles. 
Hence, we will first investigate how to apply the Born-Oppenheimer formalism in our scenario, and then we will give a detailed presentation of the diagonalization of $\hat{H}_{BO}(\rr_i,\rr_a)$ for pure Rydberg 
states and for the dressed case. In the first case, we shall also include the effect of the ion trapping potential.\\
We start with the full Hamilton operator as defined in Eq. (\ref{eq:Hfull}) and rearrange terms: 
\beq
\begin{aligned}
\label{eq:app-fullH}
\hat{H}=&\hat{H}_a+\hat{H}_i+\hat{H}_{ia}+\hat{H}^{t}_a+\hat{H}_L\\
=&\hat{H}_i+\frac{\pp_a^2}{2 M} +\hat{H}_{BO}+\bar V_{SO}^{ e-c}+\bar V_{SO}^{ e-i}+\bar V_{SO}^{ e-t}+\hat{K}
\end{aligned}
\eeq
with 
\beq
\begin{aligned}
\hat H_{BO}=&\frac{\pp^2}{2 \mu}-\frac{\pp^2\pp^2}{8m_e^3 c^2}+V_{Ryd}(\rr)+V_{SO}'^{ e-c}+\hat{H}_{ia}'+\hat{H}'^{t}_a+\hat{H}_{L}.\\
\end{aligned}
\eeq
Here we split the spin-orbit terms of the electron 
\beq
V_{SO}^{ e-\circ}=\frac{1}{2 m_e^2 c^2 }\, \bs\cdot(\;\circ\;\times\pp_e)=V_{SO}'^{ e-\circ}+\bar V_{SO}^{ e-\circ}
\eeq
into two parts 
\beq
\label{eq:splits}
V_{SO}'^{ e-\circ}=\frac{1}{2 m_e^2 c^2 }\, \bs\cdot(\;\circ\;\times\pp) \;\;\;\;\;\; \text{and} \;\;\;\;\;\; \bar V_{SO}^{ e-\circ}=\frac{m_e}{M}\frac{1}{2 m_e^2 c^2 }\, \bs\cdot(\;\circ\;\times\pp_a),
\eeq
since $\pp_e=\pp+\frac{m_e}{M}\pp_a$ in the coordinate frame we introduced in Sec.~\ref{subsec:BO}. Accordingly the $\hat{H}_{\circ}'$ are defined as $\hat{H}_{\circ}$ with $V_{SO}^{e-\circ}$ replaced by $V_{SO}'^{e-\circ}$.
The operator $\hat{K}$ comprises the terms of the relativistic kinetic energy correction, that also include the momentum operator $\pp_a$ of the atom 
\beq
\hat{K}=-\frac{m_e}{M}\frac{1}{8m_e^3 c^2}\left[4\pp^2 \pp\cdot\pp_a-4\frac{m_e}{M}\left(\left(\pp\cdot\pp_a\right)^2+2\pp^2\pp_a^2\right)-4\left(\frac{m_e}{M}\right)^2\pp\cdot\pp_a \pp_a^2-\left(\frac{m_e}{M}\right)^3\pp_a^2\pp_a^2\right].
\eeq
The so defined Born-Oppenheimer Hamiltonian, $\hat{H}_{BO}$, commutes with $\rr_i$ and $\rr_a$, and therefore we can express $\hat{H}_{BO}$ as $\int\int\mathrm{d}\rr_i \mathrm{d}\rr_a\ket{\rr_i,\rr_a}\bra{\rr_i,\rr_a}\otimes\hat{H}_{BO}(\rr_i,\rr_a)$, 
where $\hat{H}_{BO}(\rr_i,\rr_a)$ is now an operator acting on the $\rr$-Hilbert space with $\rr_i$ and $\rr_a$ treated as parameters. Hence, the full Hamiltonian~(\ref{eq:app-fullH}) becomes
\beq
\begin{aligned}
\hat{H}=&\hat{H}_i+\frac{\pp_a^2}{2 M} +\int\int\mathrm{d}\rr_i \mathrm{d}\rr_a\ket{\rr_i,\rr_a}\bra{\rr_i,\rr_a}\otimes\hat{H}_{BO}(\rr_i,\rr_a)+\bar V_{SO}^{ e-c}+\bar V_{SO}^{ e-i}+\bar V_{SO}^{ e-t}+\hat{K}.
\end{aligned}
\eeq
In our case, however, $\hat{H}_{BO}(\rr_i,\rr_a)$ is time dependent, because of the external electric fields used to trap both the atom and the ion [see Eqs.~(\ref{eq:Hion}) and~(\ref{eq:trap-atom})].
Nevertheless, as we shall see later in the appendix, we will perform a unitary transformation to $\hat{H}_{BO}(\rr_i,\rr_a)$ such that the resulting Hamiltonian can be approximated by a time-independent one. Given this, we shall now consider $\hat{H}_{BO}(\rr_i,\rr_a)$ as time independent. \\
The first step is to change to a spectral representation of $\hat{H}_{BO}(\rr_i,\rr_a)$ for each tuple of parameters $(\rr_i,\rr_a)$
\beq
\hat{H}_{BO}(\rr_i,\rr_a)=\sum_k\epsilon_k(\rr_i,\rr_a)\ket{\phi_k(\rr_i,\rr_a)}\bra{\phi_k(\rr_i,\rr_a)},
\eeq
where $\phi_k(\rr_i,\rr_a,\rr)$ denote the eigenstates of $\hat{H}_{BO}(\rr_i,\rr_a)$ with eigenenergies $\epsilon_k(\rr_i,\rr_a)$.
We assume that we can index the eigenstates such that for $\int\int\mathrm{d}\rr_i \mathrm{d}\rr_a\ket{\rr_i,\rr_a}\bra{\rr_i,\rr_a}\otimes\ket{\phi_k(\rr_i,\rr_a)}=\ket{\phi_k}$ with $k$ fixed the projectors $\hat P_k$ defined by
\beq
(\hat P_k \Psi)(\rr_i,\rr_a,\rr)=\phi_k(\rr_i,\rr_a,\rr)\int \mathrm{d}\rr'\phi^*_k(\rr_i,\rr_a,\rr') \Psi(\rr_i,\rr_a,\rr')=f_k(\rr_i,\rr_a)\phi_k(\rr_i,\rr_a,\rr)
\eeq
do exist.
To ease notation, let us define a new tensor product structure $\tilde{\otimes}$ given by $\bra{\rr_i,\rr_a,\rr}(\ket{f}\tilde{\otimes}\ket{\phi_k})=f(\rr_i,\rr_a)\phi_k(\rr_i,\rr_a,\rr)$.
Since $\sum_k \hat P_k$ adds up to identity and $\hat P_l \hat P_k$ equals $\delta_{lk} \hat P_k$, because $\phi_k(\rr_i,\rr_a,\rr)$ form a orthonormal basis of the $\rr$-Hilbert space for every tuple $(\rr_i,\rr_a)$, we can write the full Hamiltonian as:

\beq
\begin{aligned}
\hat{H}=&\sum_{l,k}\hat P_l\hat{H}\hat P_k\\
=&\sum_{k}\hat{H}_{eff}^{(k)}\tilde{\otimes}\ket{\phi_k}\bra{\phi_k}\\
&+\sum_{l,k}\hat P_l\left(\bar V_{SO}^{ e-c}+\bar V_{SO}^{ e-i}+\bar V_{SO}^{ e-t}+\hat{K}\right)\hat P_k\\
&+\sum_{l,k}\left(\hat{h}_{l,k}+\frac{1}{2 m_i}\hat{\mathbf{h}}^{(i)}_{l,k}(\rr_i,\rr_a)\cdot\pp_i+\frac{1}{2 M}\hat{\mathbf{h}}^{(a)}_{l,k}(\rr_i,\rr_a)\cdot\pp_a\right)\tilde{\otimes}\ket{\phi_l}\bra{\phi_k},\\
\end{aligned}
\eeq
where $\hat{H}_{eff}^{(k)}=\hat{H}_i+\frac{\pp_a^2}{2 M}+\epsilon_k(\rr_i,\rr_a)$ is acting on the $f_k$ component of the wave function only [cf. Eq. (\ref{eq:H_eff})] and the terms in the last line are defined as:
\beq
\begin{aligned}
\hat{h}_{l,k}(\rr_i,\rr_a)=&\int d\rr'\phi^*_l(\rr_i,\rr_a,\rr')\left(\left(\frac{\pp_i^2}{2 m_i}+\frac{\pp_a^2}{2 M}\right)\phi_k(\rr_i,\rr_a,\rr')\right),\\  
\hat{\mathbf{h}}^{(i)}_{l,k}(\rr_i,\rr_a)=&\int\mathrm{ d}\rr'\phi^*_l(\rr_i,\rr_a,\rr')\left(\pp_i\phi_k(\rr_i,\rr_a,\rr')\right),\\
\hat{\mathbf{h}}^{(a)}_{l,k}=&\int\mathrm{ d}\rr'\phi^*_l(\rr_i,\rr_a,\rr')\left(\pp_a\phi_k(\rr_i,\rr_a,\rr')\right).
\end{aligned}
\eeq
One can see this by looking at the action of $\hat H$ on a general state $\vert\Psi\rangle$ in coordinate representation:
\beq
\label{eq:BOneg}
\begin{aligned}
(\hat{H}\Psi)(\rr_i,\rr_a,\rr)=&\sum_{l,k}\left(P_l\hat{H}P_k\Psi\right)(\rr_i,\rr_a,\rr)\\
=&\sum_k\phi_k(\rr_i,\rr_a,\rr)\left(\hat{H}_i
+\frac{\pp_a^2}{2 M}
+\epsilon_k(\rr_i,\rr_a)\right)\int d\rr'\phi^*_k(\rr_i,\rr_a,\rr') \Psi(\rr_i,\rr_a,\rr')\\
&+ \left(\sum_{l,k}P_l\left(\bar V_{SO}^{ e-c}+\bar V_{SO}^{ e-i}+\bar V_{SO}^{ e-t}+\hat{K}\right)P_k\Psi\right)(\rr_i,\rr_a,\rr)\\
&+\sum_{l,k}\phi_l(\rr_i,\rr_a,\rr')\int\mathrm{ d}\rr'\phi^*_l(\rr_i,\rr_a,\rr')\left(\left(\frac{\pp_i^2}{2 m_i}
+\frac{\pp_a^2}{2 M}\right)\phi_k(\rr_i,\rr_a,\rr')\right)\int \mathrm{d}\rr''\phi^*_k(\rr_i,\rr_a,\rr'') \Psi(\rr_i,\rr_a,\rr'')\\
&+\sum_{l,k}\frac{1}{2 m_i}\phi_l(\rr_i,\rr_a,\rr')\int\mathrm{ d}\rr'\phi^*_l(\rr_i,\rr_a,\rr')\left(\pp_i\phi_k(\rr_i,\rr_a,\rr')\right)\left(\pp_i\int \mathrm{d}\rr''\phi^*_k(\rr_i,\rr_a,\rr'') \Psi(\rr_i,\rr_a,\rr'')\right)\\
&+\sum_{l,k}\frac{1}{2 M}\phi_l(\rr_i,\rr_a,\rr')\int\mathrm{ d}\rr'\phi^*_l(\rr_i,\rr_a,\rr')\left(\pp_a\phi_k(\rr_i,\rr_a,\rr')\right)\left(\pp_a\int \mathrm{d}\rr''\phi^*_k(\rr_i,\rr_a,\rr'') \Psi(\rr_i,\rr_a,\rr'')\right).\\
\end{aligned}
\eeq
We assume that we can choose $\epsilon_k(\rr_i,\rr_a)$ and $\phi_k(\rr_i,\rr_a,\rr)$ in such a way that they are differentiable in $(\rr_i,\rr_a)$ for fixed $k$. In our case this is guaranteed in a natural way, since we first project on a finite-dimensional subspace 
of the $\rr$ coordinates such that finite dimensional perturbation theory for small changes in $\rr_i$ and $\rr_a$ can be applied. Then, the second term is of order $\mathcal{O}(\frac{m_e}{M})$ [see Eq. (\ref{eq:splits})] and it can be neglected. We also neglect the last three
terms, which are small for the range of $(\rr_i,\rr_a)$ we consider, since the Eigenstates $\phi_k(\rr_i,\rr_a,\rr)$ depend just weakly on $(\rr_i,\rr_a)$. Hence, we can approximate $\hat{H}$ by an orthogonal sum of effective Hamiltonians $\hat H \simeq \bigoplus_{k}\left[\hat{H}_{\rm eff}^{(k)}\tilde{\otimes}\ket{\phi_k}\bra{\phi_k}\right]$, where
\beq
\label{eq:effham}
\hat{H}_{\rm eff}^{(k)}=
\hat{H}_i
+\frac{\pp_a^2}{2 M}
+\epsilon_k(\rr_i,\rr_a)
\eeq
each acting on the $f_k(\rr_i,\rr_a)$ component of $\vert\Psi\rangle$ only. We note that the effect of the neglected terms are corrections to the operators $\hat{H}_{\rm eff}^{(k)}$ as well as couplings between the orthogonal subspaces the 
$\hat{H}_{\rm eff}^{(k)}$ act on.\\

Now we continue in analyzing the time dependent Born-Oppenheimer Hamiltonian. Therefore we have to investigate the time dependent Schr\"odinger equation $i\hbar \partial_t \Psi=\hat{H}\Psi$. 
We want to move to a rotating frame such that the Born-Oppenheimer Hamiltonian can be written in a form such that we can apply the rotating wave approximation. To this end, we make the following ansatz for the time dependent unitary transformation:
\beq
\begin{aligned}
\hat{\mathbf{U}}=\int\int\mathrm{d}\rr_i \mathrm{d}\rr_a\ket{\rr_i,\rr_a}\bra{\rr_i,\rr_a}\otimes \hat U_{\rr_i,\rr_a}(t).
\end{aligned}
\eeq
Thus, the Schr\"odinger equation is now equivalent to $i\hbar \partial_t \hat{\mathbf{U}}\vert\Psi\rangle=\left( \hat{\mathbf{U}}\hat{H}\hat{\mathbf{U}}^\dagger+i\hbar \hat{\mathbf{U}}'\hat{\mathbf{U}}^\dagger\right)\hat{\mathbf{U}}\vert\Psi\rangle$ with $\hat{\mathbf{U}}'$ being the time 
derivative of $\hat{\mathbf{U}}$.
Let us have a closer look to the right-hand side of the Schr\"odinger equation:
\beq
\begin{aligned}
\label{eq:A12}
 \hat{\mathbf{U}}\hat{H}\hat{\mathbf{U}}^\dagger+i\hbar \hat{\mathbf{U}}'\hat{\mathbf{U}}^\dagger=&
 \hat{\mathbf{U}}\left(\hat{H}_i+\frac{\pp_a^2}{2 M}\right)\hat{\mathbf{U}}^\dagger\\
& +\int\int\mathrm{d}\rr_i \mathrm{d}\rr_a\ket{\rr_i,\rr_a}\bra{\rr_i,\rr_a}\otimes\left( \hat U_{\rr_i,\rr_a}\hat H_{BO}(\rr_i,\rr_a) \hat U_{\rr_i,\rr_a}^\dagger+i\hbar \hat U'_{\rr_i,\rr_a}\hat U_{\rr_i,\rr_a}^\dagger\right) \\
&+\hat{\mathbf{U}}\left(\bar V_{SO}^{ e-c}+\bar V_{SO}^{ e-i}+\bar V_{SO}^{ e-t}+\hat{K}\right)\hat{\mathbf{U}}^\dagger\\
=&\hat{H}_i+\frac{\pp_a^2}{2 M}\\
& +\int\int\mathrm{d}\rr_i \mathrm{d}\rr_a\ket{\rr_i,\rr_a}\bra{\rr_i,\rr_a}\otimes \left( \hat U_{\rr_i,\rr_a}\hat H_{BO}(\rr_i,\rr_a) \hat U_{\rr_i,\rr_a}^\dagger+i\hbar \hat U'_{\rr_i,\rr_a}\hat U_{\rr_i,\rr_a}^\dagger\right) \\
&+\left[\hat{\mathbf{U}},\left(\frac{\pp_i^2}{2 m_i}+\frac{\pp_a^2}{2 M}\right)\right]\hat{\mathbf{U}}^\dagger+\hat{\mathbf{U}}\left(\bar V_{SO}^{ e-c}+\bar V_{SO}^{ e-i}+\bar V_{SO}^{ e-t}+\hat{K}\right)\hat{\mathbf{U}}^\dagger\\
\approx&\hat{H}_i+\frac{\pp_a^2}{2 M}\\
& +\int\int\mathrm{d}\rr_i \mathrm{d}\rr_a\ket{\rr_i,\rr_a}\bra{\rr_i,\rr_a}\otimes \hat{\tilde{H}}_{BO}(\rr_i,\rr_a)\\
&+\left[\hat{\mathbf{U}},\left(\frac{\pp_i^2}{2 m_i}+\frac{\pp_a^2}{2 M}\right)\right]\hat{\mathbf{U}}^\dagger+\hat{\mathbf{U}}\left(\bar V_{SO}^{ e-c}+\bar V_{SO}^{ e-i}+\bar V_{SO}^{ e-t}+\hat{K}\right)\hat{\mathbf{U}}^\dagger\\
=&\sum_{{k}}\left(\hat{H}_i
+\frac{\pp_a^2}{2 M}
+\tilde{\epsilon}_k(\rr_i,\rr_a)\right)\tilde{\otimes}\ket{\tilde{\phi}_{{k}}}\bra{\tilde{\phi}_{{k}}}\\
&+\sum_{l,k}\left(\hat{\tilde{h}}_{l,k}+\frac{1}{2 m_i}\hat{\tilde{\mathbf{h}}}^{(i)}_{l,k}(\rr_i,\rr_a)\cdot\pp_i+\frac{1}{2 M}\hat{\tilde{\mathbf{h}}}^{(a)}_{l,k}(\rr_i,\rr_a)\cdot\pp_a\right)\tilde{\otimes}\ket{\phi_l}\bra{\phi_k}\\
&+\left[\hat{\mathbf{U}},\left(\frac{\pp_i^2}{2 m_i}+\frac{\pp_a^2}{2 M}\right)\right]\hat{\mathbf{U}}^\dagger\\
&+\hat{\mathbf{U}}\left(\bar V_{SO}^{ e-c}+\bar V_{SO}^{ e-i}+\bar V_{SO}^{ e-t}+\hat{K}\right)\hat{\mathbf{U}}^\dagger.
\end{aligned}
\eeq
Here we assume that we can choose $\hat U_{\rr_i,\rr_a}$ such that $ \left( \hat U_{\rr_i,\rr_a}\hat H_{BO}(\rr_i,\rr_a) \hat U_{\rr_i,\rr_a}^\dagger+i\hbar \hat U'_{\rr_i,\rr_a}\hat U_{\rr_i,\rr_a}^\dagger\right)\approx\hat{\tilde{H}}_{BO}(\rr_i,\rr_a)$, where $\hat{\tilde{H}}_{BO}(\rr_i,\rr_a)$ is a time independent operator. The $\tilde{P}_k$, $\tilde{\epsilon}_k(\rr_i,\rr_a)$, $\ket{\tilde{\phi}_{k}}$, $\hat{\tilde{h}}_{ l, k}$, $\hat{\tilde{\mathbf{h}}}_{ l, k}^{(i)}$ and $\hat{\tilde{\mathbf{h}}}_{ l, k}^{(a)}$ are defined as before, but for $\hat{\tilde{H}}_{BO}(\rr_i,\rr_a)$. As in Eq. (\ref{eq:BOneg}), the last term in Eq.~(\ref{eq:A12}) is of order $\mathcal{O}(\frac{m_e}{M})$, and the terms in the two lines above are in our case both of the type of the last term in Eq.~(\ref{eq:BOneg}). Hence, we are in a similar situation as in the time independent case discussed at the beginning of this section. We shall see in the next section how $\hat U_{\rr_i,\rr_a}$ is chosen in practice.

\subsection{Diagonalization of the Born-Oppenheimer Hamiltonian}\label{sec_Diagonalisation}
Now let us have a more detailed look at how to get the desired spectral representations of the Born-Oppenheimer Hamiltonian

\beq
\begin{aligned}
\hat H_{BO}(\mathbf{r}_i,\rr_a)=&\frac{\pp^2}{2 \mu}+V_{Ryd}(\rr)-\frac{\pp^2\pp^2}{8m_e^3 c^2}+V_{SO}'^{ e-c}\\
&-\frac{e^2}{4\pi \epsilon_0 |\RR+\frac{m_c}{M}\rr|}
+\frac{e^2}{4\pi \epsilon_0 |\RR-\frac{m_e}{M}\rr|}
+V_{SO}'^{ e-i}
+\hat{H}'^{t}_a+\hat{H}_{L}.\\
\end{aligned}
\eeq
We are interested in the case of atom-ion distances $R$ in the $\mu$m range with the electron close to the core. In the close to core region, the potential $V_{Ryd}$ dominates the remaining potential terms. Therefore we assume that projecting on a subspace of bound states of a single atom $\hat{H}_0=\frac{\pp^2}{2 \mu}+V_{Ryd}(\rr)-\frac{\pp^2\pp^2}{8m_e^3 c^2}+V_{SO}'^{ e-c}$ yields a good approximation.  To simplify the situation further we expand the ion part $\hat{H}_{ia}'$ in $\rr/R$, since $\rr/R$ is small for the states we project on, yielding the following 
approximated Hamiltonian\\
\beq
\label{eq:atomionpert}
\begin{aligned}
\hat{H}_{ia}'=&-\frac{e^2}{4\pi \epsilon_0 |\RR+\frac{m_c}{M}\rr|}
+\frac{e^2}{4\pi \epsilon_0 |\RR-\frac{m_e}{M}\rr|}
-\frac{1}{2 m_e^2 c^2 }\, \bs\cdot\left(\left(\frac{e^2 (\RR+\frac{m_c}{M}\rr)}{4\pi \epsilon_0 |\RR+\frac{m_c}{m_c+m_e}\rr|^3}\right)\times\pp\right)\\
\approx&\frac{e^2}{4\pi \epsilon_0}\left(\frac{-\rr\cdot\RR+\frac{m_c-m_e}{2M}\rr^2}{|\RR|^3}-\frac{3(m_c-m_e)(\rr\cdot\RR)^2}{2M|\RR|^5} \right)
-\frac{e^2}{8 \pi \epsilon_0 m_e^2 c^2 }\, \bs\cdot\left(
\left(\frac{\frac{m_c}{M}\rr-\RR}{|\RR|^3}+\frac{3 m_c}{M}\frac{\rr\cdot\RR}{|\RR|^5}\,\RR\right)\times\pp\right).\\
\end{aligned}
\eeq
We now use the identity 
\beq
\label{eq:momsubs}
\pp=i 2 \frac{\mu}{ \hbar} [\hat H_0,\rr]-\frac{\mu}{\hbar m_e^2 c^2} (\hat{\mathbf{S}}\times(\nabla_e V_{Ryd}))-i\frac{\mu}{ 4\hbar m_e^3 c^2}\left[\pp^2\pp^2,\rr\right].
\eeq
to substitute the momentum operator $\pp$ in Eq. (\ref{eq:atomionpert}). Because $\alpha^2=\left(\frac{e^2}{4\pi \epsilon_0 \hbar c}\right)^2 \lesssim\mathcal{O}(\frac{\rr}{R})$ for the distances we are interested in, we neglect terms of the form $\alpha^2\mathcal{O}(\frac{\rr^2}{R^2})$ and $\alpha^4\mathcal{O}(\frac{\rr}{R})$ in order to obtain an approximate expression of $\hat{H}_{ia}'$ up to $\mathcal{O}(\frac{\rr^2}{R^2})$:
\beq
\begin{aligned}
\hat{H}_{ia}' \approx& 
\frac{e^2}{4\pi \epsilon_0|\RR|^3}\left(-\rr\cdot\RR+\frac{m_c-m_e}{2M}\rr^2\right)-\frac{3e^2(m_c-m_e)}{8\pi\epsilon_0 M|\RR|^5}(\rr\cdot\RR)^2
-i \frac{e^2\mu}{4\pi\epsilon_0\hbar m_e^2  c^2}\frac{1}{|\RR|^3}\,\bs\cdot(\RR\times[\hat H_0,\rr]).\\
\end{aligned}
\eeq
We also use the identity (\ref{eq:momsubs}) to approximate $V_{SO}'^{ e-t}$ such that all perturbation terms are now simple polynomials of second order in the relative position variable $\rr$ with coefficients depending on the parameters $\rr_i$ and $\rr_a$. We then arrive at the following expression:
\beq
\begin{aligned}
&\begin{aligned}
\hat H_{BO}(\mathbf{r}_i,\rr_a)\approx&\phantom{(}\hat{H}_0\\
\end{aligned}\\
&
\begin{aligned}
\phantom{\hat H_{BO}(\mathbf{r}_i,\rr_a)\approx}&+\frac{e^2}{4\pi \epsilon_0|\RR|^3}\left(-\rr\cdot\RR+\frac{m_c-m_e}{2M}\rr^2\right)-\frac{3e^2(m_c-m_e)}{8\pi\epsilon_0 M|\RR|^5}(\rr\cdot\RR)^2\\
&-i \frac{e^2\mu}{4\pi\epsilon_0\hbar m_e^2  c^2}\frac{1}{|\RR|^3}\,\bs\cdot(\RR\times[\hat H_0,\rr])\\
\end{aligned}
&\left.
\begin{aligned}
&\phantom{\frac{R}{R}}\\
&\phantom{\frac{R}{R}}\\
\end{aligned}
 \right\}&\hat{H}_{ia}' \\
&
\begin{aligned}
\phantom{\hat H_{BO}(\mathbf{r}_i,\rr_a)\approx}&+e\phi_{PT}(\rr_c,t)-e\phi_{PT}(\rr_e,t)+i\frac{e \mu}{ \hbar\, m_e^2 c^2 }\, \bs\cdot(\mathbf{E}_{PT}(\rr_e,t)\times[\hat H_0,\rr])\\
\end{aligned}
&\left.
\begin{aligned}
&\phantom{\frac{R}{R}}\\
\end{aligned}\right\}&\hat{H}'^{t}_a\\
&
\begin{aligned}
\phantom{\hat H_{BO}(\mathbf{r}_i,\rr_a)\approx}&+e\rr\cdot\left(\mathbf{E}_{\rm dress}(\rr_a,t)+\mathbf{E}_{\rm dip}(\rr_a,t)\right).\\
\end{aligned}
&\left.
\begin{aligned}
&\phantom{\frac{R}{R}}\\
\end{aligned}\right\}&\hat{H}_{L}\\
\end{aligned}
\eeq
\clearpage
\begin{floatingfigure}[l]{0.45\linewidth}%
\label{fig_comparison}
  \vspace{\dimexpr0.3\baselineskip-\topskip}%
  \noindent
  \includegraphics[width=0.39\linewidth]{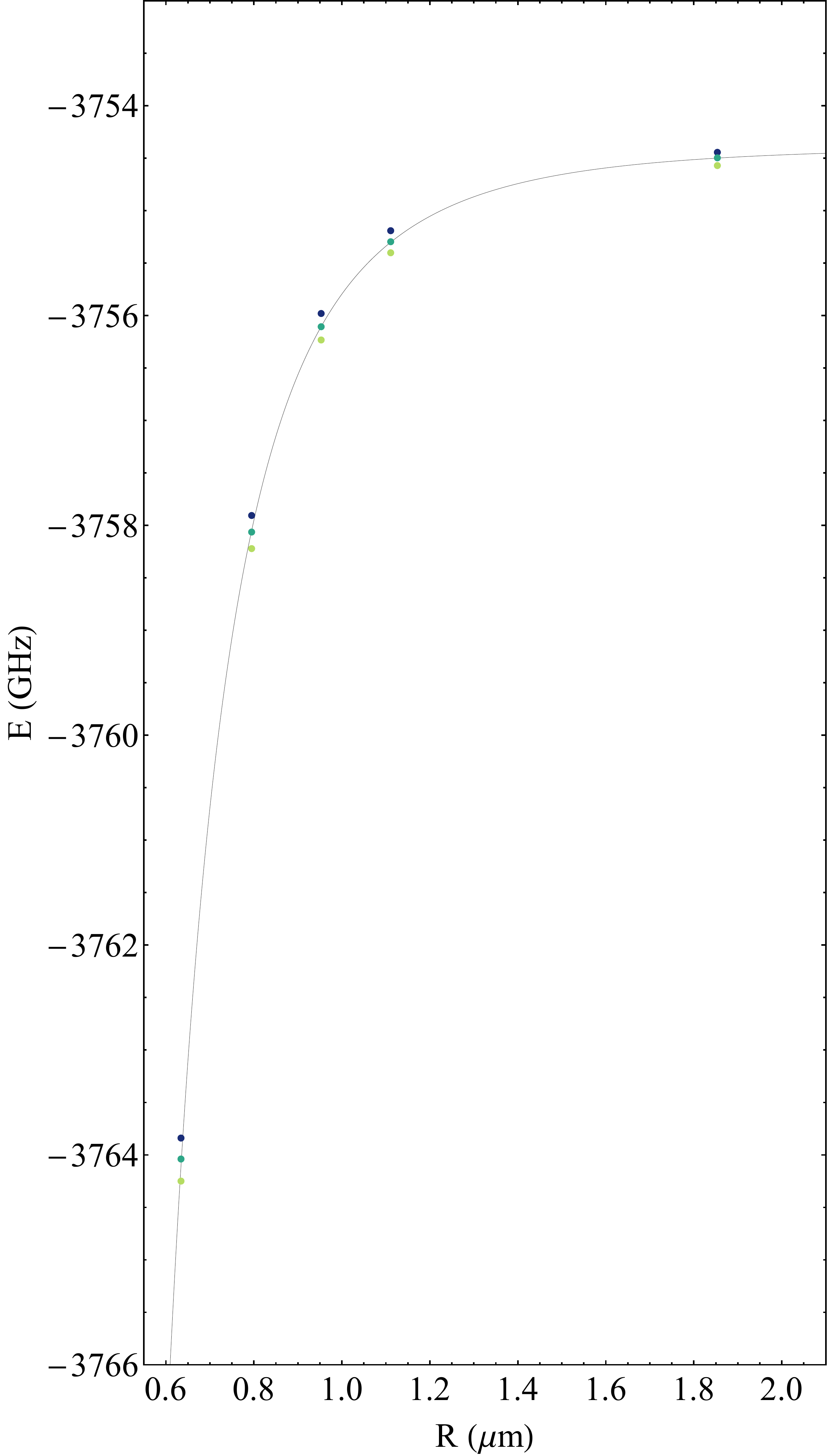}%
\caption{ Comparison of the effective potential emanating from the 30S of $^6$Li state with (dots) and without  (gray line) trapping field in the radial direction. The three points correspond to the trapping field at maximal poitive and negative and zero amplitude. In this calculation we used the following parameters, $\Omega_{rf}=2\pi~2.5$~MHz, $q=0.28$, assuming an Yb$^+$ ion trapped at the origin.}
\end{floatingfigure}
In order to proceed, we need to determine the eigenstates of $\hat H_0$. 
We assume that the effective Rydberg potential $V_{Ryd}$ depends only on the absolute 
value of $\rr$ and that for $|\rr|\rightarrow \infty$ it scales as $|\rr|^{-1}$. Under these assumptions, the angular component of the wavefunction is solved by means of the spherical 
harmonics, as for the hydrogen atom.
As far as the radial part of the wavefunction is concerned, since the exact shape of the inner part of the Rydberg potential is unknown, we 
can only rely on the experimentally determined quantum defect energy values.
Thus, we obtain the approximate radial wavefunctions by means of the Numerov 
method as in Ref.~ \cite{vanDitzhuijzen:2009}, in which we start the propagation from the classical forbidden region, where the wavefunction has to vanish at large 
$|\rr|\approx 2n(n+15) a_0$ with $a_0$ the Bohr radius and $n$ the principle quantum number, to a 
point close to the core $|\rr|\approx n^* \{n^* - [n^{*2} - (l + 1/2)^2]^{\frac1{2}}\}a_0$ by assuming a potential $\propto |\rr|^{-1}$ for all $\rr$. 
Here, $n^*=n-\delta_{nlj}$ with $\delta_{nlj}$ being the quantum defect for which we used the known 
quantum defect values reported in Ref.~\cite{Goy:1986}.\\
Let us first treat the Born Oppenheimer Hamiltonian quasi statically, which means treating time as a parameter.
In our simulations we projected on the subspace spanned by the eigenstates with principle quantum number between 25 and 35 for the case without trapping field and on the subspace spanned by the eigenstates with principle quantum number between 26 and 34, with azimuthal quantum number up to 25, for the simulations including the ion trapping field. The eigenenergies of the resulting matrices have been obtained numerically. The effect of the ion trapping field in the radial direction is visualized in Fig.~\ref{fig_comparison}, where we have chosen the field parameters as in section \ref{sec:atompaul}.\\\\
Now we continue with the time dependent case.
We can divide $\hat{H}_{BO}(\rr_i,\rr_a)$ into a time independent and a time dependent part
\beq
\begin{aligned}
\hat{H}_{BO}^{stat}(\rr_i,\rr_a)=&\frac{\pp^2}{2 \mu}+V_{Ryd}(\rr)+V_{SO}'^{ e-c}+\hat{H}_{ia}'\\
\hat{H}_{BO}^{dyn}(\rr_i,\rr_a,t)=&\hat{H}'^{t}_a+\hat{H}_{L}.\\
\end{aligned}
\eeq
\noindent
To change to a spectral representation of $\hat{H}_{BO}^{stat}(\rr_i,\rr_a)$ we use second order perturbation theory in the dipole approximation and project on a finite subspace of states, whose energy differences are close to the laser frequencies:
\beq
\hat{H}_{BO}(\rr_i,\rr_a)=\sum_k\epsilon^{stat}_k(\rr_i,\rr_a)\ket{\phi_k(\rr_i,\rr_a)}\bra{\phi_k(\rr_i,\rr_a)}+\sum_{k',k} h^{dyn}_{k',k}(\rr_i,\rr_a,t)\ket{\phi_{k'}(\rr_i,\rr_a)}\bra{\phi_k(\rr_i,\rr_a)}.
\eeq
Here $\phi_k(\rr_i,\rr_a,\rr)$ are the eigenvectors of $\hat{H}_{BO}^{stat}(\rr_i,\rr_a)$ with eigenenergies $\epsilon^{stat}_k(\rr_i,\rr_a)$, and $ h^{dyn}_{k',k}(\rr_i,\rr_a,t)=\bra{\phi_{k'}(\rr_i,\rr_a)}\hat{H}_{BO}^{dyn}(\rr_i,\rr_a,t)\ket{\phi_k(\rr_i,\rr_a)}$. For the unitary transform we use
\beq
\begin{aligned}
\hat{\mathbf{U}}=\sum_{\rr_i,\rr_a,k}\ket{\rr_i,\rr_a}\bra{\rr_i,\rr_a}\otimes e^{i \omega_k t}\ket{\phi_k(\rr_i,\rr_a)}\bra{\phi_k(\rr_i,\rr_a)},
\end{aligned}
\eeq
where the $\omega_k$ are chosen such that all the time dependence is compriced in fast rotating terms, which we neglect, namely:
\beq
\begin{aligned}
\hat{\tilde{H}}_{BO}(\rr_i,\rr_a)=& \left(\sum_{k',k}\left[\left(\epsilon^{stat}_k(\rr_i,\rr_a)-\hbar\omega_k\right)\delta_{k',k}+ e^{i (\omega_{k'}-\omega_k) t}h^{dyn}_{k',k}(\rr_i,\rr_a,t)\right]\ket{\phi_{k'}(\rr_i,\rr_a)}\bra{\phi_k(\rr_i,\rr_a)}\right) \\
\approx& \left(\sum_{k',k}\left[\left(\epsilon^{stat}_k(\rr_i,\rr_a)-\hbar\omega_k\right)\delta_{k',k}+h^{RW}_{k',k}(\rr_i,\rr_a)\right]\ket{\phi_{k'}(\rr_i,\rr_a)}\bra{\phi_k(\rr_i,\rr_a)}\right). 
\end{aligned}
\eeq

In the specific case of rydberg dressing discussed in section \ref{sec:dressing} we project on the three level subspace spanned by $\ket{g}$, $\ket{e}$ and $\ket{R}$
\begin{equation}
\hat{H}_{BO}(R)=\left( \begin{array}{ccc}
\epsilon_g &2\hbar \Omega_d(\rr_a)\cos\left(\left(\frac{\epsilon_e-\epsilon_g}{\hbar}+\Delta_d\right)t\right) &2\hbar\Omega\cos\left(\left(\frac{\epsilon_R^{(0)}-\epsilon_g}{\hbar}+\Delta_0\right)t\right) \\
2\hbar \Omega_d(\rr_a)\cos\left(\left(\frac{\epsilon_e-\epsilon_g}{\hbar}+\Delta_d\right)t\right) &\epsilon_e & 0\\
2\hbar\Omega \cos\left(\left(\frac{\epsilon_R^{(0)}-\epsilon_g}{\hbar}+\Delta_0\right)t\right)  & 0 &\epsilon_R^{(0)}-\frac{C_4^{|R\rangle}}{R^4} \end{array} \right),
\end{equation}
where we neglect the $R$ dependence of $\Omega$, $\epsilon_e$ and $\epsilon_g$. According to this we choose 
\begin{equation}
\hat U(R,t)=\left( \begin{array}{ccc}
e^{-i \frac{\epsilon_g}{\hbar}t} &0 & 0 \\
0 &e^{-i(\frac{\epsilon_e}{\hbar}+\Delta_d)t} & 0\\
0 & 0 &e^{-i(\frac{\epsilon_R^{(0)}}{\hbar}+\Delta_0)t} \end{array} \right)
\end{equation}
for the unitary transform .
The resulting three dimensional matrix is the one  given in Eq. (\ref{eq:H3level}), which can be diagonalized to get the effective potential (\ref{eqVad}) in the Rydberg dressed case.

\section{Micromotion calculation}
\label{app:mm}

For the sake of simplicity and without loss of generality, in the numerical simulations for the assessment of the impact of the micromotion on both the atom and ion we have considered 
the following total atom-ion Hamiltonian:

\begin{align}
\label{eq:Htapp}
\hat H_{tot}(t) &= \hat H_{mm}(t) + \hbar\omega_a\left(\hat b^\dag\hat b + \frac{1}{2}\right)
+ \eta\hbar\Omega_{S-M}\cos(\omega_v t)\left(\hat a^\dag 
+ \hat a \right) \ket{\uparrow}_i\bra{\uparrow} + \frac{\tilde{V}^{(3)}(\hat x_i,\hat x_a)}{2}\left[1 + \cos(\omega_v t)\right]  \ket{\uparrow}_a\bra{\uparrow},
\end{align}
where terms rotating faster than $\delta$ have not been neglected. 
The other terms in Eq.~(\ref{eq:mmHtot}) are then defined as:

\begin{align}
\hat H_{mm}(t) & = \frac{\hat p_i^2}{2 m_i} + \frac{m_i \Omega_{\rm rf}^2 q}{4} \hat x_i^2 \cos(\rf t), \nonumber
\end{align}
\begin{align}
\tilde{V}^{(3)}(\hat x_i,\hat x_a) = \tilde{V}(0,0) + \sum_{k=1}^3\tilde{V}_{0,k}^{\prime}(0,0) \hat x_a^k + \sum_{k=1}^3\tilde{V}_{k,0}^{\prime}(0,0) \hat x_i^k  + \tilde{V}_{1,1}^{\prime}(0,0) \hat x_i \hat x_a + \tilde{V}_{1,2}^{\prime}(0,0) \hat x_i \hat x_a^2 + \tilde{V}_{2,1}^{\prime}(0,0) \hat x_i^2 \hat x_a, \nonumber
\end{align}
\begin{align*}
\tilde{V}_{j,k}^{\prime}(0,0) & = \frac{1}{(j+k)!}\frac{\partial^{j+k}}{\partial x_i^j\partial x_a^k}\tilde{V}(0,0),
\end{align*}
\begin{align}
\tilde{V}(\hat x_i,\hat x_a) & = \frac{\hbar \Omega^2}{\Delta_0 + \frac{\alpha_{\ket{R}}}{2\hbar}\vert \mathbf{E}_{ion} + \mathbf{E}_{\rm s} + \mathbf{E}_{\rm rf}\vert^2}. \nonumber
\end{align}
Here the norm of the total electric field $\mathbf{E} = \mathbf{E}_{ion} + \mathbf{E}_{\rm s} + \mathbf{E}_{\rm rf}$ is given by

\begin{align}
\vert\mathbf{E}(\rr_a,\rr_i,t)\vert ^2 & = 
\frac{m_i^2\omega_i^4}{4 e^2}\left(x_a^2 + y_a^2 + 4 z_a^2\right) 
+\frac{m_i^2\rf^4 q^2}{4 e^2}\cos^2(\rf t) (x_a^2 + y_a^2) 
+ \frac{e^2 k_C^2}{\left[(x_a - x_i)^2+(y_a - y_i)^2 + (z_a - z_i)^2\right]^2}\nonumber\\
\phantom{=}&
+\frac{m_i^2\rf^2 \omega_i^2q}{2 e^2}\cos(\rf t) (x_a^2 -  y_a^2)
+m_i k_C \omega_i^2 \frac{ (x_a - x_i) x_a +  (y_a - y_i) y_a - 2 (z_a - z_i) z_a}{\left[(x_a - x_i)^2+(y_a - y_i)^2 + (z_a - z_i)^2\right]^{3/2}}\nonumber\\
\phantom{=}& 
+m_i \rf^2 q k_C\cos(\rf t) \frac{(x_a - x_i) x_a - (y_a - y_i) y_a}{\left[(x_a - x_i)^2+(y_a - y_i)^2 + (z_a - z_i)^2\right]^{3/2}}.
\end{align}
As we outlined in Sec.~\ref{sec:mm}, we consider the situation in which the atom is trapped some distance $d$ away from the ion in the (transverse) $x$ direction 
(i.e., $\bar{x}_i=0$ and $\bar{x}_a = d$) and $y_i=y_a=z_i=z_a=0$. Thus, the above expression for the electric field norm simplifies to

\begin{align}
\vert\mathbf{E}(x_a,x_i,t)\vert ^2 & = 
\frac{m_i^2}{4 e^2}(x_a+d)^2\left[
\omega_i^4 + \rf^4 q^2\cos^2(\rf t) + 2 q \omega_i^2 \rf^2 \cos(\rf t) 
\right] \nonumber\\&+\frac{e^2 k_C^2}{(x_a - x_i + d)^4}
+m_i k_C \frac{x_a+d}{(x_a - x_i +d)^2} \left[\omega_i^2 + \rf^2 q \cos(\rf t) \right].\nonumber
\end{align}
Now, by rewriting the adiabatic potential as $\tilde{V}(\hat x_i,\hat x_a) =  \frac{\xi_1}{\xi_2 + \xi_3 f(x_i,x_a)}$, the corresponding derivatives are:

\begin{align*}
\frac{\partial}{\partial x_{a,i}} \tilde{V}(0,0) & = - \frac{\xi_1\xi_3}{[\xi_2 + \xi_3 f(0,0)]^2}\frac{\partial}{\partial x_{a,i}}f(0,0),\\
\\
\frac{\partial^2}{\partial x_{a,i}^2} \tilde{V}(0,0) & = - \frac{\xi_1\xi_3}{[\xi_2 + \xi_3 f(0,0)]^2}\frac{\partial^2}{\partial x_{a,i}^2}f(0,0) + \frac{2\xi_1\xi_3^2}{[\xi_2 + \xi_3 f(0,0)]^3}\left(\frac{\partial}{\partial x_{a,i}}f(0,0)\right)^2,\\
\\
\frac{\partial^3}{\partial x_{a,i}^3} \tilde{V}(0,0) & = - \frac{\xi_1\xi_3}{[\xi_2 + \xi_3 f(0,0)]^2}\frac{\partial^3}{\partial x_{a,i}^3}f(0,0) + \frac{6\xi_1\xi_3^2}{[\xi_2 + \xi_3 f(0,0)]^3}\frac{\partial}{\partial x_{a,i}}f(0,0) \frac{\partial^2}{\partial x_{a,i}^2}f(0,0) -  \frac{6\xi_1\xi_3^3}{[\xi_2 + \xi_3 f(0,0)]^4}\left(\frac{\partial}{\partial x_{a,i}}f(0,0)\right)^3,\\
\\
\frac{\partial^2}{\partial x_a\partial x_i} \tilde{V}(0,0) & = - \frac{\xi_1\xi_3}{[\xi_2 + \xi_3 f(0,0)]^2}\frac{\partial^2}{\partial x_a\partial x_i}f(0,0) + \frac{2\xi_1\xi_3^2}{[\xi_2 + \xi_3 f(0,0)]^3}\frac{\partial}{\partial x_{a}}f(0,0)\frac{\partial}{\partial x_{i}}f(0,0),\\
\end{align*}
\begin{align*}
\frac{\partial^3}{\partial x_a\partial x_i^2} \tilde{V}(0,0) & = - \frac{\xi_1\xi_3}{[\xi_2 + \xi_3 f(0,0)]^2}\frac{\partial^3}{\partial x_a \partial x_i^2}f(0,0) + \frac{2\xi_1\xi_3^2}{[\xi_2 + \xi_3 f(0,0)]^3}\left[
\frac{\partial^2}{\partial x_i^2}f(0,0) \frac{\partial}{\partial x_a}f(0,0)+ 2 \frac{\partial}{\partial x_i}f(0,0) \frac{\partial^2}{\partial x_a \partial x_i}f(0,0)
\right]\\
\phantom{=} & -  \frac{6\xi_1\xi_3^3}{[\xi_2 + \xi_3 f(0,0)]^4}\left(\frac{\partial}{\partial x_{i}}f(0,0)\right)^2 \!\!\! \frac{\partial}{\partial x_{a}}f(0,0),
\end{align*}
\begin{align*}
\frac{\partial^3}{\partial x_a^2\partial x_i} \tilde{V}(0,0) & = - \frac{\xi_1\xi_3}{[\xi_2 + \xi_3 f(0,0)]^2}\frac{\partial^3}{\partial x_a^2 \partial x_i}f(0,0) + \frac{2\xi_1\xi_3^2}{[\xi_2 + \xi_3 f(0,0)]^3}\left[
\frac{\partial^2}{\partial x_a^2}f(0,0) \frac{\partial}{\partial x_i}f(0,0) + 2 \frac{\partial}{\partial x_a}f(0,0) \frac{\partial^2}{\partial x_a \partial x_i}f(0,0)
\right]\\
\phantom{=} & -  \frac{6\xi_1\xi_3^3}{[\xi_2 + \xi_3 f(0,0)]^4}\left(\frac{\partial}{\partial x_{a}}f(0,0)\right)^2 \!\!\!\frac{\partial}{\partial x_{i}}f(0,0).
\end{align*}
Here, $\xi_1 = \hbar \Omega^2$, $\xi_2 = \Delta_0$, and $\xi_3 = \frac{\alpha_{\ket{r}}}{2\hbar}$. Then, the total electric field norm is

\begin{align*}
f(0,0) &= 
\frac{m_i^2\omega_i^4}{4 e^2} d^2 
+\frac{m_i^2\rf^4 q^2}{4 e^2}\cos^2(\rf t) d^2 
+ \frac{e^2 k_C^2}{d^4}
+\frac{m_i^2\rf^2 \omega_i^2q}{2 e^2}\cos(\rf t) d^2
+ \frac{m_i k_C \omega_i^2}{d}
+ \frac{m_i \rf^2 q k_C\cos(\rf t)}{d} 
\end{align*}
while the derivatives of $f$ are:

\begin{align*}
\frac{\partial}{\partial x_a}f(0,0) &= \frac{m_i^2 d}{2 e^2}\left[
\omega_i^4 + \rf^4 q^2 \cos^2(\rf t) + 2 q \omega_i^2 \rf^2 \cos(\rf t)
\right]
-\frac{m_i k_C}{d^2}\left[
\omega_i^2 + \rf^2 q \cos(\rf t)
\right] 
-\frac{4 e^2 k_C^2}{d^5}
\frac{\partial^2}{\partial x_a^2}\\
f(0,0) &= \frac{m_i^2}{2 e^2}\left[
\omega_i^4 + \rf^4 q^2 \cos^2(\rf t) + 2 q \omega_i^2 \rf^2 \cos(\rf t)
\right]+\frac{2 m_i k_C}{d^3}\left[
\omega_i^2 + \rf^2 q \cos(\rf t)
\right] 
+\frac{20 e^2 k_C^2}{d^6}
\end{align*}
\begin{align*}
\frac{\partial^3}{\partial x_a^3}f(0,0) &= 
-\frac{6 m_i k_C}{d^4}\left[
\omega_i^2 + \rf^2 q \cos(\rf t)
\right] 
-\frac{120 e^2 k_C^2}{d^7} 
\end{align*}
\begin{align*}
\frac{\partial}{\partial x_i}f(0,0) &= 
\frac{2 m_i k_C}{d^2}\left[
\omega_i^2 + \rf^2 q \cos(\rf t)
\right] 
+\frac{4 e^2 k_C^2}{d^5}
\end{align*}
\begin{align*}
\frac{\partial^2}{\partial x_i^2}f(0,0) &= 
\frac{6 m_i k_C}{d^3}\left[
\omega_i^2 + \rf^2 q \cos(\rf t)
\right] 
+\frac{20 e^2 k_C^2}{d^6}
\end{align*}
\begin{align*}
\frac{\partial^3}{\partial x_i^3}f(0,0) &= 
\frac{24 m_i k_C}{d^4}\left[
\omega_i^2 + \rf^2 q \cos(\rf t)
\right] 
+\frac{120 e^2 k_C^2}{d^7}
\end{align*}
\begin{align*}
\!\!\!\!\frac{\partial^2}{\partial x_a \partial x_i}f(0,0) &= 
-\frac{4 m_i k_C}{d^3}\left[
\omega_i^2 + \rf^2 q \cos(\rf t)
\right] 
-\frac{20 e^2 k_C^2}{d^6}
\end{align*}
\begin{align*}
\!\!\!\!\!\!\!\!\frac{\partial^3}{\partial x_a \partial x_i^2}f(0,0) &= 
-\frac{18 m_i k_C}{d^4}\left[
\omega_i^2 + \rf^2 q \cos(\rf t)
\right] 
-\frac{120 e^2 k_C^2}{d^7}\\
\end{align*}
\begin{align*}
\!\!\!\!\frac{\partial^3}{\partial x_a^2 \partial x_i}f(0,0) &= 
\frac{12 m_i k_C}{d^4}\left[
\omega_i^2 + \rf^2 q \cos(\rf t)
\right] 
+\frac{120 e^2 k_C^2}{d^7}
\end{align*}

Finally, we note that the micromotion Hamiltonian can be rewritten as:

\begin{align}
\hat H_{mm}(t) & = \frac{\hat p_i^2}{2 m_i} + \frac{m_i\omega_i^2}{2}\hat x_i^2 + \frac{m_i \Omega_{\rm rf}^2 q}{4} \hat x_i^2 \cos(\rf t) - \frac{m_i\omega_i^2}{2}\hat x_i^2 \nonumber\\
\phantom{=} & 
= \hat H_0^{i} + \frac{m_i\omega_i^2}{2}\hat x_i^2 \left[
\frac{\rf^2}{2\omega_i^2} q  \cos(\rf t)  -1
\right].\nonumber
\end{align}
Now, the goal is to solve the Schr\"odinger equation $i\hbar\partial_t\ket{\psi(t)} = \hat H_{tot}(t)\ket{\psi(t)}$ with initial condition at time $t=0$ given by the (Gaussian) ground states 
(note that this does not exactly correspond to the ground state of the ion in the Paul trap) of the unperturbated ion $\hat H_0^{i}$ and atom $\hat H_0^{a} = \hbar\omega_{a} (n_{a} + 1/2)$ 
Hamiltonians. The equation can be then easily solved in coordinated space $(x_i,x_a)$ and the integration can be performed with a split-step operator together with the fast Fourier transform 
techniques.  

For the sake of completeness, we briefly note that for the numerics it is better to work in dimensionless units. To this end, we have rescaled the energies in units of $\hbar\rf$ and 
the lengths in units of $\ell= \sqrt{\hbar/(\mu_{ai}\bar{\omega})}$ with $\mu_{ai} = m_a m_i/(m_a+m_i)$ being the reduced mass, and $\bar{\omega} = \sqrt{\omega_a\omega_i}$. Thus, we 
shall also make the replacement $\tau =\rf t$. Then, the dressed potential can be rewritten as:

\begin{align}
\frac{\tilde{V}(\hat x_i,\hat x_a)}{\hbar\rf} = \frac{\frac{\Omega}{\rf}}{\frac{\Delta_0}{\Omega} + \frac{\alpha_{\ket{R}}}{2\hbar\Omega}\mathcal{E}^2f(\hat x_i,\hat x_a)} = 
\frac{\bar{\xi}_1}{\bar{\xi}_2 + \bar{\xi}_3 f(\hat x_i,\hat x_a)},
\end{align}
where $\bar{\xi}_1= \Omega/\rf$, $\bar{\xi}_2=\Delta_0/\Omega$, and 

\begin{align}
\bar{\xi}_3 =  \frac{\alpha_{\ket{R}}}{2\hbar\Omega}\mathcal{E}^2 = \frac{\gamma \alpha_{\ket{\uparrow}}}{2\hbar\Omega} \frac{k_C^2 e^2}{\ell^4} = 
\frac{\gamma}{4} \frac{(\hbar\bar{\omega})^2}{\hbar\Omega E^*},
\end{align}
with $E^*=\hbar^4/(2\alpha_{\ket{\uparrow}}\mu^2 e^2 k_C^2)$ and $\gamma = \alpha_{\ket{R}}/\alpha_{\ket{\uparrow}}$. Hence, we have

\begin{align}
f(0,0) &= 
\beta_1\frac{\bar{d}^2}{4} + \beta_2\frac{\bar{d}^2}{4} q^2\cos^2(\tau) + \beta_3\frac{\bar{d}^2}{2} q\cos(\tau) + \frac{\beta_4}{\bar{d}}+ \frac{\beta_5}{\bar{d}} q\cos(\tau) 
+ \frac{1}{\bar{d}^4},
\end{align}
where $\bar{d} = d/\ell$, and

\begin{align*}
\beta_1 &= \frac{m_i^2\omega_i^4\ell^6}{e^4 k_C^2}, \qquad \beta_2 = \frac{m_i^2\rf^4\ell^6}{e^4 k_C^2}, \qquad \beta_3 = \frac{m_i^2\omega_i^2\rf^2\ell^6}{e^4 k_C^2}, \qquad \beta_4 = \frac{m_i\omega_i^2\ell^3}{k_C e^2}, \qquad \beta_5 = \frac{m_i\rf^2\ell^3}{k_C e^2}.
\end{align*}

%
%

\section{Determination of the effective Rabi and driving frequencies}

Once the driving frequency $\omega_v$ and the ion bicromatic Rabi frequency $\Omega_{S-M}$ are chosen, we have to choose the Rydberg effective Rabi frequency $\Omega$ such that 
when both the atom and the ion are in the internal spin state $\vert\uparrow\rangle$ the corresponding forces compensate each other, as we discussed in Sec.~\ref{sec:spin-spin}. 
In other words, in order to have $\langle\hat x_{i}(t)\rangle \simeq 0\,\forall t$ the laser Hamiltonian $\hat H_{S-M}$ and the linear term of $\tilde{V}^{(3)}(\hat x_i,\hat x_a)$ have 
to be equal in size, but opposite in sign. More precisely, 

\begin{align}
\label{eq:condition}
2\eta\hbar\Omega_{S-M} = - V^{\prime}_{1,0}(0,0).
\end{align}
However, the derivative $V^{\prime}_{1,0}(0,0)$ is time-dependent, because of the rf-field. Since the rf-field oscillates at a frequency much higher than the observation time 
($\sim$ 1 ms), we can take the time average $\langle V^{\prime}_{1,0}(0,0) \rangle_{T_{rf}}$ with $T_{rf} = 2\pi/\rf$, that is, 

\begin{align}
\langle V^{\prime}_{1,0}(0,0) \rangle_{T_{rf}} & = \left\langle \frac{\partial}{\partial x_{i}} \tilde{V}(0,0) \right\rangle_{T_{rf}}  = -  \left\langle \frac{\xi_1\xi_3}{[\xi_2+\xi_3 f(0,0)]^2}\frac{\partial}{\partial x_{i}}f(0,0) \right\rangle_{T_{rf}}. \nonumber
\end{align}
For the sake of simplicity, we shall compute such time average as

\begin{align}
\label{eq:time-ave}
\langle V^{\prime}_{1,0}(0,0) \rangle_{T_{rf}} =  -  \frac{\xi_1\xi_3}{[\xi_2+\xi_3 \left\langle f(0,0)\right\rangle_{T_{rf}}]^2}\left\langle \frac{\partial}{\partial x_{i}}f(0,0) \right\rangle_{T_{rf}}, 
\end{align}
with 

\begin{align*}
\langle f(0,0) \rangle_{T_{rf}} = \frac{m_i^2\omega_i^4}{4 e^2} d^2 
+\frac{m_i^2\rf^4 q^2}{8 e^2}d^2 
+ \frac{e^2 k_C^2}{d^4}
+ \frac{m_i k_C \omega_i^2}{d},
\end{align*}
and

\begin{align*}
\left\langle \frac{\partial}{\partial x_{i}}f(0,0) \right\rangle_{T_{rf}} = 
\frac{2 m_i k_C}{d^2}\omega_i^2  
+\frac{4 e^2 k_C^2}{d^5}.\\
\end{align*}
Putting everything together, one has to choose the laser strength such that Eq.~(\ref{eq:condition}) is fulfilled. Given the results displayed in Fig.~\ref{fig_gate_mm}, 
we can also conclude that the computation of the time average as done in Eq.~(\ref{eq:time-ave}) is a very good approximation for the considered numerical example. 

Finally, we describe how $\omega_v$ can be chosen for particular ion trap parameters. The secular frequency of the ion can be approximated by 
$\omega_i^{(\perp)}\approx \frac{\Omega_{\rm rf}}{2}\sqrt{a+q^2/2}$ for small $q$ and $a$. For the parameters used in section~\ref{sec:mm}, where we set $a=0$, since 
we neglected the static trapping field, we get $\omega_i^{(\perp)}\approx 2\pi$~250~kHz. However, a more accurate calculation based on continued fractions for solving 
the Mathieu equations~\cite{Leibfried:2003} yields: $\omega_i^{(\perp)}= 2\pi$~254.089~kHz such that $\delta^{(\perp)} = \omega_v - \omega_i^{(\perp)} = 2\pi$~1.064~kHz in 
Sec.~\ref{sec:mm}, corresponding to a gate time of $\tau_g =$~940~$\mu$s. This in turn gives $J\tau_g/\hbar=\pi/4$, corresponding to the desired phase gate.

\twocolumngrid


%

\end{document}